\pdfoutput=1

\documentclass[journal=jctcce,manuscript=article]{achemso}

\usepackage[page]{appendix}
\usepackage[version=4]{mhchem}
\usepackage[T1]{fontenc} 
\usepackage{amsmath}
\usepackage{xfrac}
\usepackage{booktabs}
\usepackage[en-US]{datetime2}
\usepackage{notes2bib}
\usepackage{siunitx}
\DeclareSIUnit[number-unit-product = {\,}]
\cal{cal}
\DeclareSIUnit\kcal{\kilo\cal}
\usepackage{glossaries}
\newglossaryentry{Ktar}{
    name=\ensuremath{K_{\text{target}}},
    description={N/A}}
\newglossaryentry{nosehoover}{
    name=Nos{\'e}-Hoover,
    description={N/A}}
\newacronym{md}{MD}{molecular dynamics}
\newacronym{com}{COM}{center of mass}
\newacronym{pbc}{PBC}{periodic boundary conditions}
\newacronym{msd}{MSD}{mean-squared displacement}
\newacronym{mof}{MOF}{metal-organic framework}
\newacronym{rdf}{RDF}{radial distribution function}
\newacronym[longplural={degrees of freedom}]{dof}{DOF}{degree of freedom}
\newacronym{csvr}{CSVR}{canonical sampling through velocity rescaling}

\newcommand{\Lagr}{\mathcal{L}}
\newcommand{\orderof}{\mathcal{O}}

\author{Efrem Braun}
\affiliation[Berkeley]
{Department of Chemical and Biomolecular Engineering, University of California, Berkeley, Berkeley, CA 94720, USA}
\author{S.\ Mohamad Moosavi}
\affiliation[EPFL]
{Institut des Sciences et Ing{\'e}nierie Chimiques (ISIC), Valais, {\'E}cole Polytechnique F{\'e}d{\'e}rale de Lausanne (EPFL), Rue de l'Industrie 17, CH-1951 Sion, Switzerland}
\author{Berend Smit}
\email{berend.smit@epfl.ch}
\affiliation[Berkeley]
{Department of Chemical and Biomolecular Engineering, University of California, Berkeley, Berkeley, CA 94720, USA}
\alsoaffiliation[EPFL]
{Institut des Sciences et Ing{\'e}nierie Chimiques (ISIC), Valais, {\'E}cole Polytechnique F{\'e}d{\'e}rale de Lausanne (EPFL), Rue de l'Industrie 17, CH-1951 Sion, Switzerland}

\title
  {Anomalous effects of velocity rescaling algorithms: the flying ice cube
  effect revisited}

\begin{document}
\begin{singlespace}

%
%
%
%
%

\begin{abstract}
  The flying ice cube effect is a molecular dynamics simulation artifact
  in which the use of velocity rescaling thermostats sometimes causes
  the violation of the equipartition theorem, affecting both
  structural and dynamic properties.
  The reason for this artifact and the conditions under which it occurs
  have not been fully understood.
  Since the flying ice cube effect was first demonstrated, a new velocity
  rescaling algorithm (the \acrshort{csvr} thermostat) has been developed
  and become popular without its effects on the equipartition theorem
  being truly known.
  Meanwhile, use of the simple velocity rescaling and Berendsen
  thermostat algorithms has not abated but has actually continued to
  grow.
  Here, we have calculated the partitioning of the kinetic energy between
  translational, rotational, and vibrational modes in simulations of
  diatomic molecules to explicitly determine whether the equipartition
  theorem is violated under different thermostats and while rescaling
  velocities to different kinetic energy distributions.
  We have found that the underlying cause of the flying ice cube effect is
  a violation of balance leading to systematic redistributions of kinetic
  energy under simple velocity rescaling and the Berendsen thermostat.
  When velocities are instead rescaled to the canonical ensemble's
  kinetic energy distribution, as is done with the \acrshort{csvr}
  thermostat, the equipartition theorem is not violated, and we show that
  the \acrshort{csvr} thermostat satisfies detailed balance.
  The critical necessity for molecular dynamics practitioners to abandon
  the use of popular yet incorrect velocity rescaling algorithms is
  underscored with an example demonstrating that the main result of a
  highly-cited study is entirely due to artifacts resulting from the
  study's use of the Berendsen thermostat.
\end{abstract}

\glsresetall

\section{Introduction}

By integrating the classical Newtonian equations of motion, \gls{md}
simulations naturally sample the microcanonical (NVE) ensemble due to
conservation laws.\cite{fre011,lei151} For comparison with experiment, it is
often desirable to sample constant-temperature ensembles such as the
canonical (NVT) or isothermal-isobaric (NPT) ensembles. In analogy with
experiment, these ensembles could be generated by sampling a subspace of
a much larger microcanonical system that serves as a heat bath, but such
an approach is usually too computationally-expensive to implement in
practice.  Instead, various thermostatting algorithms are typically
applied to change the Hamiltonian dynamics in a manner such that the
intended ensemble is sampled. Many such algorithms have been proposed,
and some of the more well-known choices include:

\begin{itemize}

  \item Simple velocity rescaling, pioneered by \citet{woo711}
  for thermal equilibration, rescales the velocities of all
  particles at the end of each timestep (it can also be conducted
  with a less frequent time rescaling period) by a factor $\lambda$ to
  achieve a target instantaneous temperature: $\lambda =
  \left(\frac{{\gls{Ktar}}}{K}\right)^{\frac{1}{2}}$ with
  $\gls{Ktar}= \frac{1}{2}N_{\text{DOF}}k_{\text{B}}T_{\text{target}}$,
  where $N_{\text{DOF}}$ is the number of \acrlongpl{dof} in the system.

  \item The Gaussian thermostat supplements Newton's second law with a force
  intended to keep the kinetic energy constant:\cite{eva831,nos842,eva902}
  $\mathbf{\dot{p}}_i = -\nabla U_i - \alpha \mathbf{p}_i$, where $\alpha$
  is a Lagrange multiplier determined using Gauss' principle of least
  constraint to be $\alpha = \left. \left(\sum_{i=1}^{N} \mathbf{F}_i \cdot
  \mathbf{p}_i /m_i\right) \middle/ \left(\sum_{i=1}^{N}
  \mathbf{p}_i^2/m_i\right) \right.$.

  \item Langevin dynamics supplements Newton's second law with terms describing
  Brownian motion:\cite{sch781} $\mathbf{\dot{p}}_i = -\nabla U_i - \gamma
  \mathbf{p}_i + \mathbf{\eta}$, where $\gamma$ represents a frictional
  dissipative force and $\mathbf{\eta}(t,T,\gamma,m_i)$ is a stochastic
  term representing random collisions.

  \item The Berendsen thermostat takes the Langevin equation, removes
  the stochastic term, and modifies the frictional dissipative force
  to yield similar temperature time dependence as with the
  stochastic term present:\cite{ber841} $\mathbf{\dot{p}}_i =
  -\nabla U_i - \gamma \mathbf{p}_i \left( \frac{\gls{Ktar}}{K} -1
  \right)$, where $\gls{Ktar}=
  \frac{1}{2}N_{\text{DOF}}k_{\text{B}}T_{\text{target}}$.  In
  practice, this is implemented as a smoother version of the simple
  velocity rescaling technique, in which the velocities of all
  particles are rescaled at the end of each timestep by a factor
  $\lambda$, with $\lambda=\left[1+\frac{\Delta
  t}{\tau_T}\left(\frac{\gls{Ktar}}{K}-1\right)\right]^{\frac{1}{2}}$.
  $\tau_T$ represents a time damping constant; if it is set equal to
  the timestep, the Berendsen algorithm recovers simple velocity
  rescaling, and as the time damping constant approaches infinity,
  the Berendsen algorithm recovers conventional microcanonical
  dynamics.

  \item The \gls{csvr} thermostat is a velocity rescaling algorithm in which the
  velocities of all particles are rescaled at the end of each timestep by
  a factor $\lambda$ designed such that the kinetic energy exhibits the
  distribution of the canonical ensemble.\cite{bus071,hey831} To this end,
  $\lambda = \left(\frac{{\gls{Ktar}}}{K}\right)^{\frac{1}{2}}$, where
  $\gls{Ktar}$ is stochastically drawn from the probability density
  function $P(\gls{Ktar}) \propto \gls{Ktar}^{\sfrac{N_{\text{DOF}}}{2}-1} e^{-\beta
  \gls{Ktar}}$.  This algorithm can be adjusted to yield a smoother
  evolution in a similar manner as the Berendsen algorithm smoothes simple
  velocity rescaling.\cite{bus071}

  \item The \gls{nosehoover} thermostat extends the classical Lagrangian to
  include the additional coordinate $s$ and its
  time-derivative:\cite{nos841,hoo851} $\Lagr = s^2\sum_{i=1}^N
  \frac{\mathbf{p}_i^2}{2m_i} - U + \frac{1}{2}Q\dot{s}^2 -
  k_{\text{B}}T_{\text{target}}L\ln s$, where $Q$ is the effective
  ``mass'' associated with $s$ and $L$ is set by the number of \acrlongpl{dof}. A single
  \gls{nosehoover} thermostat may be used, or chains of thermostats  may
  be implemented to improve ergodicity and to take into account additional
  conservation laws.\cite{mar921}

\end{itemize}

There exist numerous additional thermostats (e.g., the Andersen
thermostat\cite{and801}), and small changes can be made to the listed
thermostats, such as implementing the originally global \gls{nosehoover}
thermostat in a local ``massive'' manner by pairing a separate
\gls{nosehoover} thermostat to each \acrlong{dof}.\cite{tob931} The reader
is referred to a non-comprehensive list of reviews and textbooks for
additional information.\cite{mor981,hun051,fre011,tuc101}

Simple velocity rescaling and the Gaussian thermostat aim to sample the
isokinetic ensemble (NVK). However, they are often presented as
equivalent to the canonical ensemble with respect to position-dependent
equilibrium properties, with justification for this based on the
argument that the configurational part of the isokinetic ensemble's
partition function is exactly equal to that of the canonical
ensemble's.\cite{hai831,eva832,nos842,min031,col101}
Meanwhile, the
Berendsen thermostat does not correspond to a known ensemble but is
rather supposed to sample a configurational phase space intermediate to
the canonical and microcanonical ensembles.\cite{ber841,mor001,mor031}

In the 1990s, it was found that the simple velocity rescaling and
Berendsen thermostat algorithms introduce an
artifact:\cite{lem941,har981} the ``flying ice cube effect,'' as coined
by \citet{har981}, describes a violation of the equipartition theorem
observed when using these algorithms in which kinetic energy drains from
high-frequency modes such as bond stretching into low-frequency modes
such as \gls{com} translation. This was shown to affect systems'
structural, thermodynamic, and dynamic properties.\cite{har981}
As it can be proven that the equipartition theorem
holds in the canonical ensemble, microcanonical ensemble, and isokinetic
ensemble (see Appendix),\cite{cal851,cag881,shi061,uli081,sib131}
a simulation exhibiting the flying ice cube effect is not ergodically
sampling any of these ensembles, neither in configurational phase space
nor in momentum phase space.

\begin{figure}
  \centering
  \includegraphics[width=3.46in]{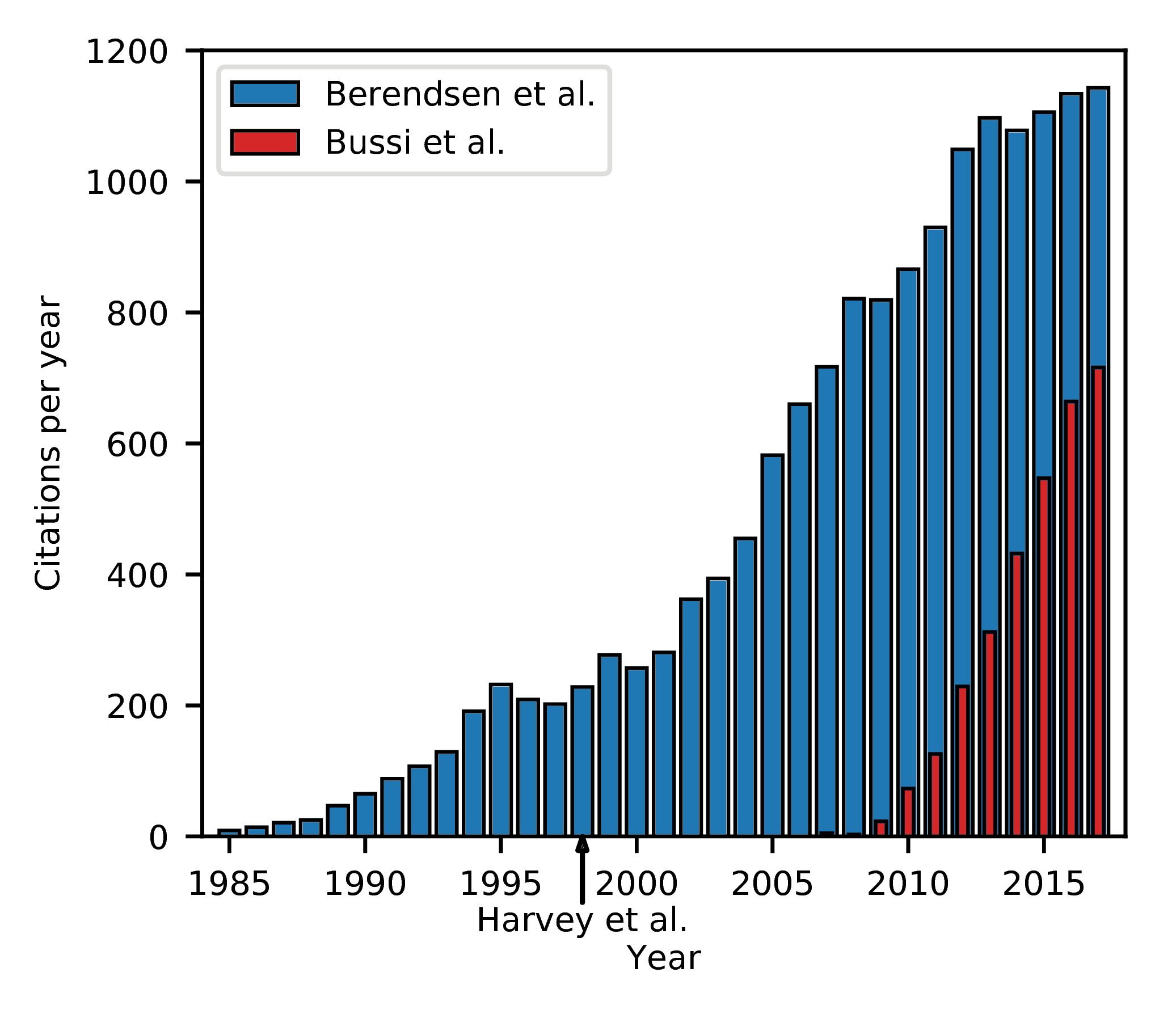}
  \caption{\label{fig:thermostatcitations}
    Citations of \citet{ber841} and \citet{bus071} over time. Data
    provided by Web of Science, extracted on \DTMdate{2018-05-04}.}
\end{figure}

Nonetheless, simple velocity rescaling and the Berendsen thermostat
continue to be commonly used,\cite{hun051,coo081} with \citet{coo081}
stating, ``By far the most commonly used algorithm for constant
temperature MD of biomolecules is the Berendsen heat bath, due to its
ease of implementation and availability in standard software packages.''
Use of the Berendsen thermostat can be approximated by tracking
citations of its canonical reference,\cite{ber841} which have continued
to grow over time (Fig.~\ref{fig:thermostatcitations}).

Some technical aspects of the flying ice cube effect are as of yet still
unclear.  Since \citet{har981}, there has been continued discussion
about whether the flying ice cube effect may occur with other
thermostats.\cite{lin081,gog121} The \gls{csvr} thermostat rescales
velocities to yield the canonical ensemble's distribution of kinetic
energies, similar to how simple velocity scaling yields the isokinetic
ensemble's distribution of kinetic energies and the Berendsen thermostat
yields a kinetic energy distribution intermediate to the two ensembles.
If all velocity rescaling algorithms always lead to the flying ice cube
effect, then it may be suspected that the same flying ice cube artifact
occurs when using the \gls{csvr} thermostat,\cite{bas131} which would be
worrisome because the \gls{csvr} thermostat has been quickly adopted
into widespread use (Fig.~\ref{fig:thermostatcitations}).  In addition,
since the Gaussian thermostat has been shown to be similar to simple
velocity rescaling,\cite{nos911} it may be suspected that the Gaussian
thermostat exhibits the artifact as well.  Given the wide-spread use of
these algorithms in \gls{md} simulations,
more understanding is warranted, and we will show that neither the
\gls{csvr} thermostat nor the Gaussian thermostat bring about the flying
ice cube effect.

In the present work we refer to the flying ice cube effect as the term
was originally used to describe the violation of the equipartition
theorem as caused by velocity rescaling procedures.\cite{har981} Other
\gls{md} simulation methods that fail to conserve energy in the
microcanonical ensemble can also bring about equipartition theorem
violations.\cite{lin081} These methods include approximate treatment of
long-range electrostatic interactions, certain multiple timestep
algorithms, constraining molecular geometries with too loose of a
tolerance, not updating neighbor lists frequently enough, and using too
large of a timestep.\cite{chi001,lin081,eas101} In some cases these
issues are also referred to as flying ice cube
effects,\cite{sag991,wag131,yan132} but these are not related to the
artifact with which we are concerned.

In this work, we have revisited the simple model system of united-atom
diatomic ethane molecules that \citet{har981} first used to illustrate
the flying ice cube effect.  By explicitly calculating the partitioning
of kinetic energies between translational, rotational, and vibrational
\acrlongpl{dof}, we are able to determine which thermostats and conditions
lead to the violation of equipartition, as well as the manner and degree
to which they do so.
We go on to rationalize these findings by illustrating how simple
velocity rescaling violates balance, while the \gls{csvr} thermostat
satisfies detailed balance.
We end by illustrating some severe errors that are directly caused by
these subtleties related to thermostatting.

\section{Simulation Details}

Diatomic ethane molecule simulations were conducted with the open-source
LAMMPS code.\cite{pli951}
LAMMPS input scripts are available.
\bibnote{We used the \DTMdate{2016-11-17} release of LAMMPS to conduct
our simulations. The Gaussian thermostat was not implemented in
LAMMPS, so we wrote an extension that integrates the equations of
motion given by \citet{min031}.  This extension was later
incorporated into the LAMMPS code and made publicly available
starting with the \DTMdate{2017-01-06} update as part of the ``fix nvk''
command.}

Except where stated otherwise, the simulations consisted of cubic
simulation boxes with \gls{pbc},
setup by placing the ethane molecules on a simple cubic lattice,
equilibrated with a Langevin thermostat for at least
\SI{50}{\nano\second},
switched to the target thermostat for at least a further
\SI{50}{\nano\second} of equilibration, and finally ran with the target
thermostat for at least \SI{50}{\nano\second} of production.
We verified that all simulations were conducted for sufficient time
periods for the energies to equilibrate and be well sampled.  The
velocities of the particles in microcanonical simulations were rescaled
once after Langevin equilibration such that the total energy was equal
to the average total energy seen in the Langevin simulation.  For the
simulations in which the \gls{com} linear momentum was fixed to zero
(stated in the figure captions), the system's linear momentum was zeroed
every timestep, followed by a rescaling of velocities to maintain the
same total kinetic energy as before the zeroing had occurred to prevent
energy leakage. The equations of motion were integrated with a standard
Velocity Verlet algorithm using half-step velocity calculations.  The
timestep used was \SI{0.5}{\femto\second},
which was found to give adequate energy conservation in the
microcanonical ensemble.

Thermostat parameters were as follows, except where stated otherwise.
Simple velocity rescaling was done every timestep. The \gls{nosehoover}
chain consisted of three thermostats. The Berendsen, \gls{nosehoover},
and \gls{csvr} thermostats were used with time damping constants
($\tau_T$) of \SI{100}{\femto\second}, and the \gls{nosehoover}
thermostat used effective thermostat masses of $Q_1=N_{\text{DOF}}
k_{\text{B}} T {\tau_T}^2$ and $Q_{i>1}=k_{\text{B}} T
{\tau_T}^2$.\cite{mar921} When doing simulations in the microcanonical
ensemble, the total energy was set such that a simulation
temperature equal to the canonical ensemble simulations' target
temperature was achieved.
The target simulation temperature was set to \SI{350}{\kelvin}, well
above the critical temperature of ethane.\cite{mar982}

Kinetic energies of each diatomic molecule were partitioned into
translational, rotational, and vibrational kinetic energies, as shown in
the Appendix.  In all figures that plot kinetic
energies, the error bars shown represent $\pm1$ standard error of the
mean. This was calculated by dividing the data from the production
timesteps into \num{20} consecutive blocks, averaging the data for each
block, and computing the standard error over the \num{20} data
values.\cite{fre011} Error bars are not shown when they would be smaller
than the symbols or the line widths.

Bonded parameters for the united-atom ethane molecule were taken from
\citet{har981} (harmonic bond potential $U(r)=k(r-r_0)^2$ with
$r_0=$\SI{1.54}{\angstrom} and
$k=$\SI{240}{\kcal\per\mol\per\angstrom\squared}) and non-bonded
parameters were taken from \citet{mar982} (Lennard-Jones potential with
$\epsilon=$\SI{0.195}{\kcal\per\mole}, $\sigma=$\SI{3.75}{\angstrom},
truncated and shifted at \SI{14}{\angstrom}, and no charges).

Details on the simulations of benzene in \acrshort{mof}-5 can be found
in the Appendix.

\section{Results and Discussion}

\subsection{Examining equipartition under different thermostats}

\begin{figure}
  \centering
  \includegraphics[width=3.32in]{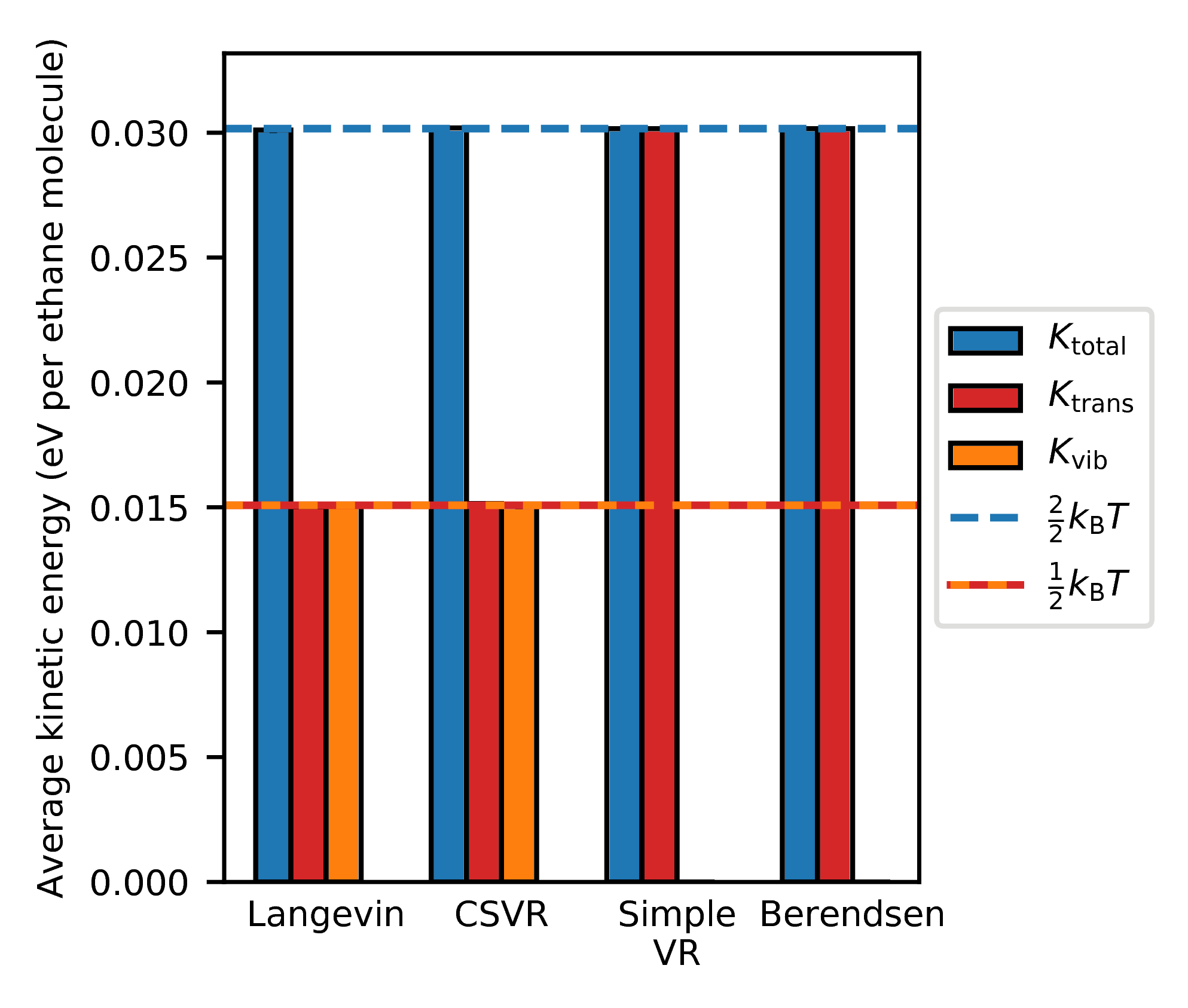}
  \caption{\label{fig:1partsequipartition}
    Partitioning of the kinetic
    energies obtained from one-dimensional \gls{md} simulations of a
    single ethane molecule using various thermostats.  Both atoms were
    given a starting velocity of \SI{100}{\meter\per\second} along the
    same direction as the bond vector. For the thermostats shown, the
    same energy partitionings were observed regardless of initial bond
    length and initial \gls{com} momentum.  The microcanonical,
    \gls{nosehoover} thermostat, and Gaussian thermostat results are not
    shown here since we found that the energy partitionings are
    dependent on the initial conditions, indicative of these thermostats'
    well-known lacks of ergodicity that are more manifest for small
    systems.\cite{fre011,nos842,tox901,nos911,mar921,tuc011,hes031,hun051}}
\end{figure}

It is instructive to reconsider the simple case previously examined by
\citet{har981}: that of a single ethane molecule moving in
one-dimensional space along its bond axis. In the microcanonical
ensemble under perfect energy conservation, the translational kinetic
energy will remain constant at its set initial energy and the
vibrational kinetic energy will oscillate.
In the canonical ensemble, equipartition states that the translational
and vibrational \acrlong{dof} should each have an average kinetic energy of
$\frac{1}{2}k_{\text{B}} T$.  As expected, the Langevin thermostat
satisfies the equipartition theorem (see
Fig.~\ref{fig:1partsequipartition}). In agreement with the work of
\citet{har981}, we find that simple velocity rescaling and the Berendsen
thermostat bring about a violation of equipartition in the kinetic
\acrlongpl{dof}, with all kinetic energy flowing to translational motion, in
the plainest illustration of the flying ice cube effect.
We find that the \gls{csvr} thermostat correctly partitions the
energies.

\begin{figure}
  \centering
  \includegraphics[width=6.87in]{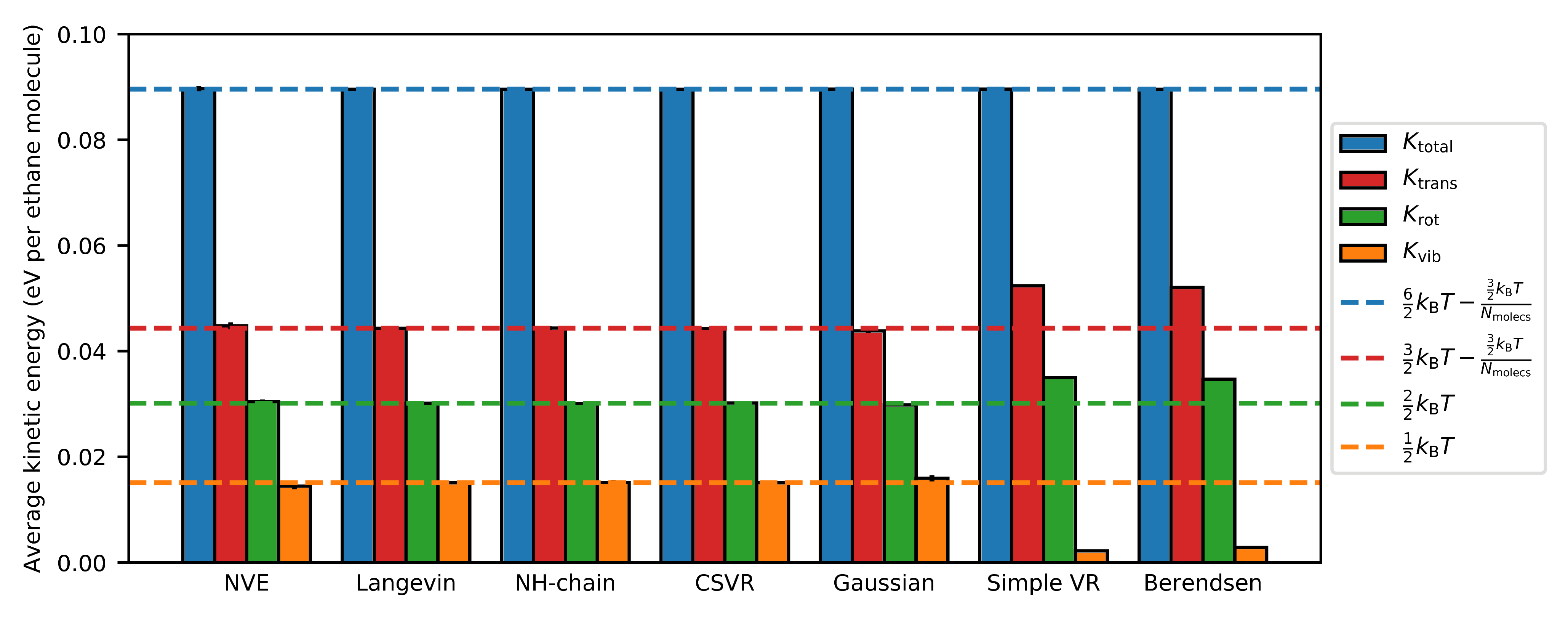}
  \caption{\label{fig:50partsequipartition}
    Partitioning of the kinetic energies obtained from \gls{md}
    simulations of 50 ethane molecules in a \SI{30}{\angstrom} cubic
    simulation box using various thermostats. In all simulations shown,
    the \gls{com} momentum was fixed to zero.}
\end{figure}

We next consider the more complex case of a large number of ethane
molecules interacting in three dimensions with anharmonic Lennard-Jones
potentials. Each diatomic ethane molecule now has three translational
modes, two rotational modes, and one vibrational modes, so the
equipartition theorem states that these modes' kinetic energies should
be equal to $\frac{3}{2}k_{\text{B}} T$, $\frac{2}{2}k_{\text{B}} T$,
and $\frac{1}{2}k_{\text{B}} T$ respectively, with a correction of
$\frac{3}{2}k_{\text{B}} T/N_{\rm{molecs}}$ to the translational
kinetic energy in cases where the \gls{com} momentum is constrained.  In
Fig.~\ref{fig:50partsequipartition}, we show that the Langevin,
\gls{nosehoover}, \gls{csvr}, and Gaussian thermostats all exhibit
correctly equipartitioned energies, as does the microcanonical ensemble.
As in the case of the single ethane molecule in one dimension, the
simple velocity rescaling and Berendsen thermostat algorithms lead to a
violation of equipartition, with translational and rotational modes
having too much kinetic energy and vibrational modes having too little.

\subsection{Equivalence of simple velocity rescaling and the Gaussian
thermostat}

Since the thermostatting under simple velocity rescaling does not take
place within the equations of motion, this ad hoc temperature control
algorithm was initially difficult to investigate theoretically, and its
validity was considered questionable.\cite{nos842,eva902} The
algorithm's use was justified on the basis of empirical arguments, such
as that simple velocity rescaling and the Gaussian thermostat give
similar static and dynamic properties for the Lennard-Jones
fluid.\cite{hai831} It was eventually proven that simple velocity
rescaling is analytically equivalent to the Gaussian thermostat within
an error of $\orderof \left(\text{timestep}\right)$ when the velocity
rescaling time period is set equal to the timestep,\cite{nos911} which
gave support for the legitimacy of using simple velocity rescaling to
sample the isokinetic ensemble.

However, we have shown that the Gaussian thermostat exhibits correct
energy equipartitioning while simple velocity rescaling does not.  We
prove in the Appendix that the isokinetic ensemble should
satisfy the equipartition theorem. Thus, it is clear that simple
velocity rescaling does not actually sample the isokinetic ensemble.

\begin{figure}
  \centering
  \includegraphics[width=6.87in]{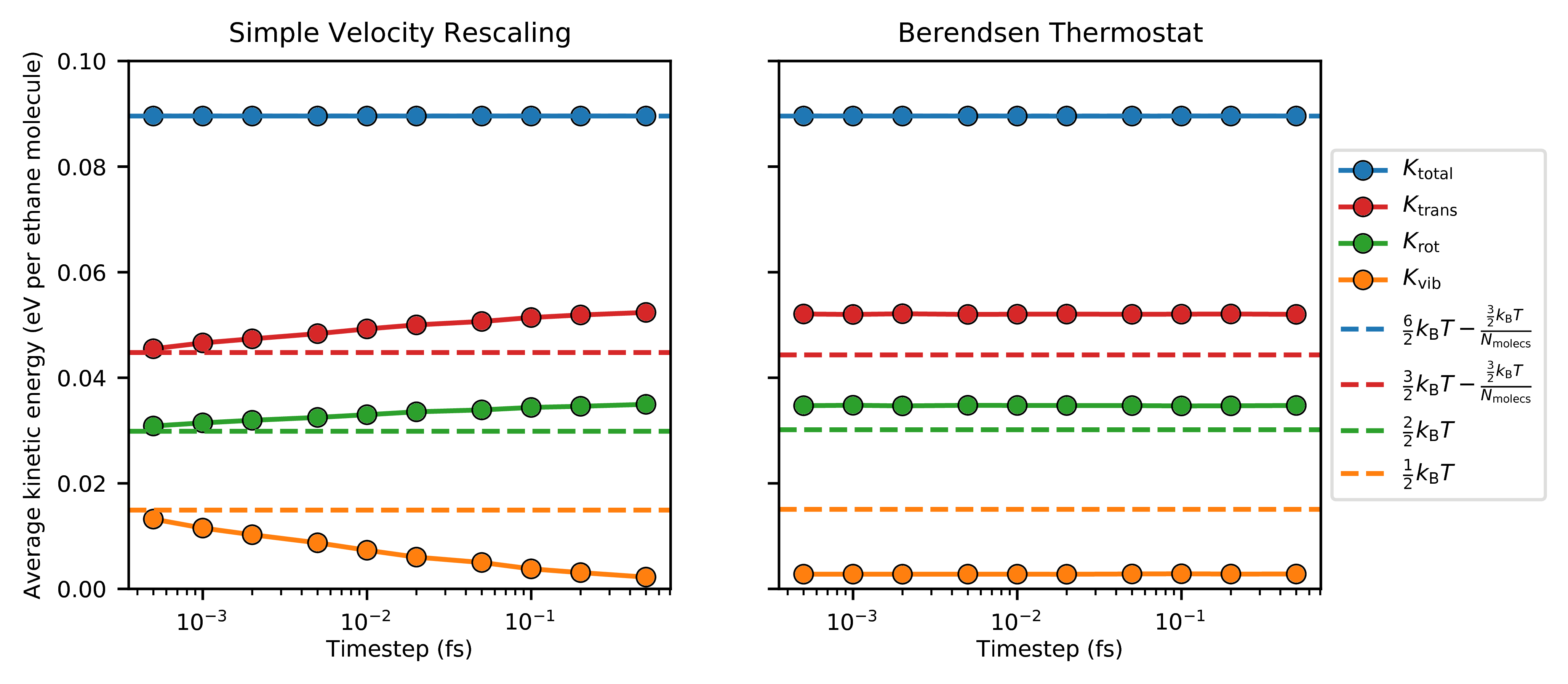}
  \caption{\label{fig:timestep}
    Partitioning of the kinetic energies obtained from \gls{md}
    simulations performed under the same conditions as in
    Fig.~\ref{fig:50partsequipartition} but changing the timestep, using
    (left) simple velocity rescaling and (right) the Berendsen
    thermostat with the time damping constant maintained at
    \SI{100}{\femto\second}. Lines are a guide to the eye.}
\end{figure}

The equivalence of simple velocity rescaling and the Gaussian thermostat
under small timesteps leads to the expectation that the flying ice cube
effect will gradually disappear under simple velocity rescaling as the
timestep is decreased.  We demonstrate confirmation of this expectation
in Fig.~\ref{fig:timestep}.  However, Fig.~\ref{fig:timestep} shows that
the timestep needs to be reduced by over three orders of magnitude from
typical simulation timesteps before the flying ice cube effect is no
longer discerned.
Of course, such a decreased timestep requires an equivalent three orders
of magnitude increase in CPU time; if the timestep between integrations
is so small, the forces on the particles should not need to be
recalculated every timestep, and so one could envision implementing a
multiple-time-step algorithm to mitigate the increase in CPU time.
We also note that under the Berendsen thermostat, lowering the timestep
does not correct the energy partitioning.

\subsection{Violation of balance causes the flying ice cube effect}

\begin{figure}
  \centering
  \includegraphics[width=3.43in]{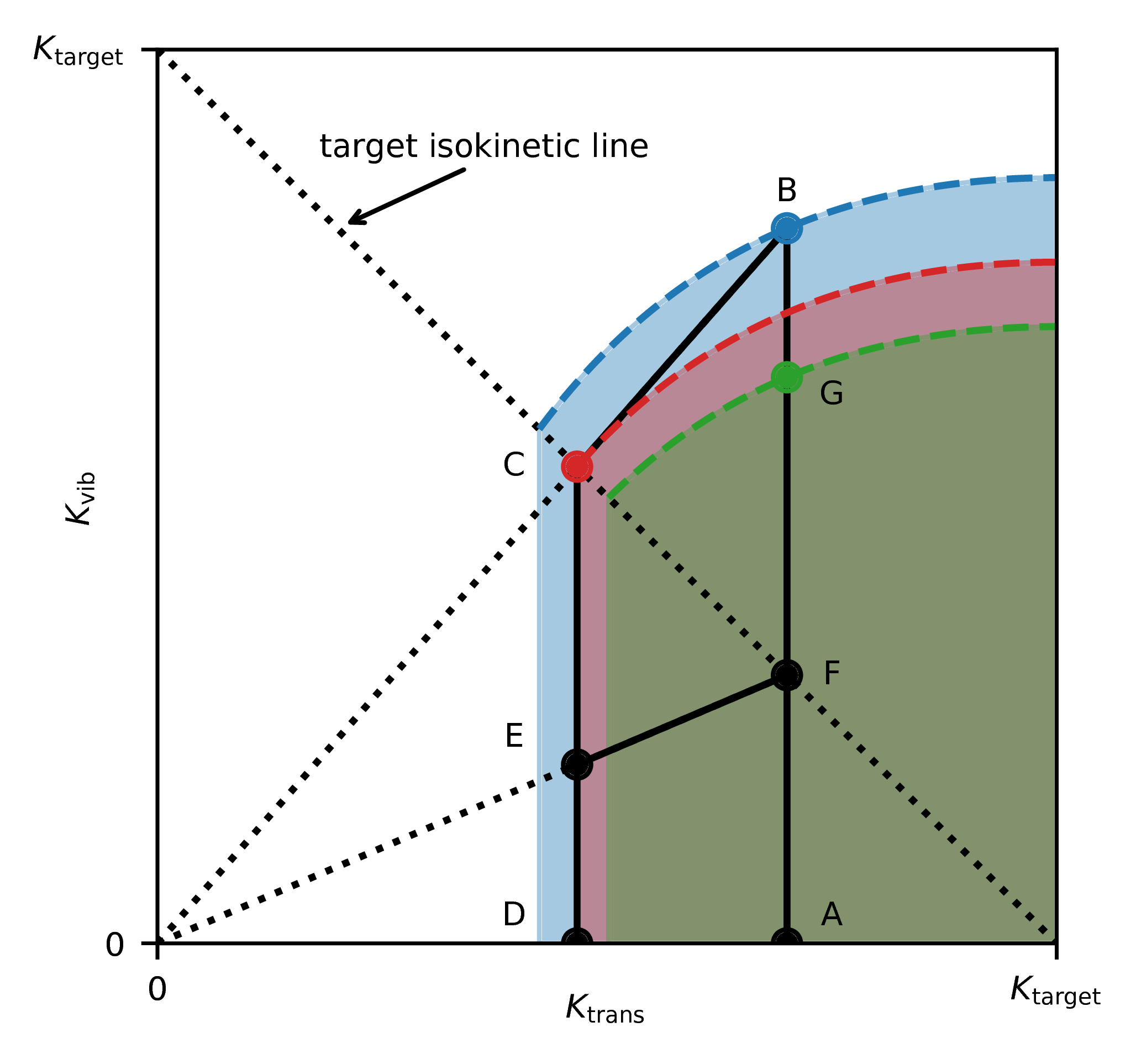}
  \caption{\label{fig:balanceviolation}
  Kinetic phase space of a single ethane molecule moving in
  one-dimensional space along its bond axis under simple velocity
  rescaling.  $\gls{Ktar}= k_{\text{B}}T_{\text{target}}$,
  $K_{\rm{trans}}=\frac{1}{2}\left(m_1+m_2\right)\left(\frac{m_1v_{1,x}+m_2v_{2,x}}{m_1+m_2}\right)^{^2}$,
  and
  $K_{\rm{vib}}=\frac{1}{2}\left(\frac{m_1m_2}{m_1+m_2}\right)\left(v_{2,x}-v_{1,x}\right)^2$.
  Solid lines show a particular path in phase space between labeled
  points, referred to in the text.  Dotted lines are guides useful to
  understanding the velocity rescaling moves.  Dashed lines show the
  boundaries of phase space accessible by any sequence of \gls{md} and
  velocity rescalings from lines $\overline{AB}$, $\overline{CD}$,
  and $\overline{AG}$, with the accessible phase spaces shaded.}
\end{figure}

The mechanism underlying the flying ice cube effect can be elucidated
graphically for the first test case we examined, that of a single ethane
molecule. In Fig.~\ref{fig:balanceviolation}, we show this system's
phase space, putting translational kinetic energy on the $x$-axis and
vibrational kinetic energy on the $y$-axis.

During microcanonical \gls{md}, the system can only explore phase space
on a vertical line between $y=0$ and $y=U_{\rm{max}}$ because a constant
total energy and translational kinetic energy is maintained, with energy
exchanges only allowed between vibrational kinetic energy and potential
energy.  Consider a \gls{md} simulation initially on such a vertical
line in phase space, $\overline{AB}$. Under simple velocity rescaling,
if a rescaling move is conducted at point $B$, the system will move to
point $C$; this occurs because the translational and vibrational
energies are both scaled by the same factor $\lambda^2$ such that their
sum is equal to the target kinetic energy, moving the system to the
intersection of the lines $y=\frac{y_B}{x_B}x$ and the target isokinetic
line ($y=-x+K_{\rm{target}}$). Since points $B$ and $C$ have the same
configuration with zero potential energy, \gls{md} will now explore line
$\overline{CD}$.

Let us examine whether we can reach point $B$ by rescaling from line
$\overline{CD}$ back to a line with the same translational energy of
line $\overline{AB}$. With a single rescaling, we would need to rescale
from point $E$ to point $F$.  From point $F$, \gls{md} will explore
phase space on line $\overline{AG}$, where the lengths of lines
$\overline{FG}$ and $\overline{CE}$ are equal, with both representing
the stored potential energy of the system prior to the rescaling.
Obviously, line $\overline{EF}$ must have a smaller slope than line
$\overline{BC}$; accordingly, $y_G$ will necessarily be smaller than
$y_B$. Hence, with a single velocity rescaling, point $B$ cannot be
reached.  Multiple velocity rescalings from line $\overline{CD}$ allows
us to reach a point with greater vibrational kinetic energy than point
$G$.  However, all phase space reachable by any number of velocity
rescalings from line $\overline{CD}$ is bounded by the red dashed line
in Fig.~\ref{fig:balanceviolation} (see Appendix for
derivation).  Continuing to rescale will continue to shrink the volume
of accessible phase space, as rescaling from lines $\overline{AB}$ to
$\overline{CD}$ to $\overline{AG}$ lowers the boundary from the blue to
the red to the green dashed lines; eventually, accessible phase space
will be confined only to the point with all kinetic energy in the
translational mode.

Notably, the decrease in accessible phase space becomes smaller as
velocity rescaling occurs closer to the isokinetic line. In a
simulation, this occurs when the timestep between velocity rescalings is
reduced.  This explains why the flying ice cube effect is reduced under
simple velocity rescaling by decreasing the timestep
(Fig.~\ref{fig:timestep}).

\subsubsection{Monte Carlo perspective}

We can view the combination of \gls{md} and velocity scaling moves as a
Monte Carlo simulation.  Hence, our previous example shows that simple
velocity rescaling violates the condition of
balance.\cite{man991,fre011}

In contrast, the \gls{csvr} thermostat can explicitly be proven to
sample the desired distribution by considering the condition of detailed
balance.  Let us assume that we do a large and random number of \gls{md}
steps between velocity rescaling moves. We define $A$ as the set of all
configurations of the system with a total energy $E_A$.  The flow of
configurations from set $A$ to set $B$ is given by:
\begin{align}
    & K\left(A \rightarrow B\right) = \nonumber \\
    & \quad P\left(E_A\right)
    \sum_{{\bf r}_1^n} \sum_{{\bf p}_1^n} \sum_{{\bf r}_2^n} \sum_{{\bf p}_2^n}
    p \left({\bf r}_1^n,{\bf p}_1^n | E_A\right) 
    \delta \left(E\left({\bf r}_1^n,{\bf p}_1^n\right)-E_A\right)
    \alpha \left({\bf r}_1^n,{\bf p}_1^n \rightarrow {\bf r}_2^n,{\bf p}_2^n\right)
    \delta \left(E\left({\bf r}_2^n,{\bf p}_2^n\right)-E_B\right) \label{eq:detailedbalance1}
\end{align}
where ${\bf r}_1^n,{\bf p}_1^n$ is the configuration with position
vector ${\bf r}_1^n$ and momentum vector ${\bf p}_1^n$, $p\left({\bf
r}_1^n,{\bf p}_1^n | E_A\right)$ is the probability to find the
configuration ${\bf r}_1^n,{\bf p}_1^n$ from all configurations with
energy $E_A$ during \gls{md},  and $\alpha \left({\bf r}_1^n,{\bf p}_1^n
\rightarrow {\bf r}_2^n,{\bf p}_2^n\right)$ is the \latin{a priori}
probability to velocity rescale from configuration ${\bf r}_1^n,{\bf
p}_1^n$ to configuration ${\bf r}_2^n,{\bf p}_2^n$.
Recognizing that velocity rescaling does not alter positions:
\begin{equation}
    K\left(A \rightarrow B\right) =
    P\left(E_A\right)
    \sum_{{\bf r}^n} \sum_{{\bf p}_1^n} \sum_{{\bf p}_2^n}
    p \left({\bf r}^n,{\bf p}_1^n | E_A\right)
    \delta \left(E\left({\bf r}_1^n,{\bf p}_1^n\right)-E_A\right)
    \alpha \left({\bf r}^n,{\bf p}_1^n \rightarrow {\bf r}^n,{\bf p}_2^n\right)
    \delta \left(E\left({\bf r}^n,{\bf p}_2^n\right)-E_B\right) \label{eq:detailedbalance2}
\end{equation}
Next, recognizing that velocity rescaling can only give one
configuration in momentum space with $E\left({\bf r}^n,{\bf
p}_2^n\right)=E_B$ from starting configuration ${\bf r}^n,{\bf p}_1^n$,
and that the acceptance probabilities only involve the kinetic energy:
\begin{equation}
    K\left(A \rightarrow B\right) =
    P\left(E_A\right)
    \sum_{{\bf r}^n} \sum_{{\bf p}^n}
    p \left({\bf r}^n,{\bf p}^n | E_A\right) 
    \delta \left(E\left({\bf r}_1^n,{\bf p}_1^n\right)-E_A\right)
    \alpha \left(K=E_A-U\left( {\bf r}^n \right) \rightarrow E_B-U\left( {\bf r}^n \right)\right) \label{eq:detailedbalance3}
\end{equation}
where $\alpha \left(K=E_A-U\left( {\bf r}^n \right) \rightarrow
E_B-U\left( {\bf r}^n \right)\right)$ is the \latin{a priori}
probability to velocity rescale to the configuration having kinetic
energy $K=E_B-U\left( {\bf r}^n \right)$ given we start with a
configuration having kinetic energy $K=E_A-U\left( {\bf r}^n \right)$.
Then, recognizing that momentum and position are decoupled, i.e., the
number of possible states in momentum space only depends on the total
kinetic energy but does not depend on the details of the potential
energy surface, and each of these possible states in momentum space are
equally likely:
\begin{equation}
    K\left(A \rightarrow B\right) =
    P\left(E_A\right)
    \sum_{{\bf r}^n}
    \omega \left( E_A-U \left( {\bf r}^n \right) \right)
    p \left({\bf r}^n,{\bf p}^n | E_A\right) 
    \alpha \left(K=E_A-U\left( {\bf r}^n \right) \rightarrow E_B-U\left( {\bf r}^n \right)\right) \label{eq:detailedbalance4}
\end{equation}
where $\omega\left( K\right)$ is the number of configurations in
momentum space for a given kinetic energy $K$ (equivalent to the ideal
gas microcanonical partition function).
Finally, by making the substitutions $p\left({\bf r}^n,{\bf p}^n |
E_A\right) = \Omega_{NVE_A}^{-1}$ and $P\left(E_A\right) =
\frac{e^{-\beta E_A} \Omega_{NVE_A}}{Z_{NVT}}$:
\begin{equation}
    K\left(A \rightarrow B\right) =
    \frac{e^{-\beta E_A}}{Z_{NVT}}
    \sum_{{\bf r}^n}
    \omega \left(E_A-U \left( {\bf r}^n \right) \right)
    \alpha \left(K=E_A-U\left( {\bf r}^n \right) \rightarrow E_B-U\left( {\bf r}^n \right)\right) \label{eq:detailedbalance5}
\end{equation}

The two flows, $K\left(A \rightarrow B\right)$ and $K\left(B \rightarrow
A\right)$, are equal if we impose as condition for the \latin{a priori}
probabilities:
\begin{align}
\frac{\alpha \left(K=E_A-U\left( {\bf r}^n \right) \rightarrow E_B-U\left( {\bf r}^n \right)\right)}
     {\alpha \left(K=E_B-U\left( {\bf r}^n \right) \rightarrow E_A-U\left( {\bf r}^n \right)\right)}
&=
\frac{e^{-\beta E_B} \omega \left( E_B-U \left( {\bf r}^n \right) \right)}
     {e^{-\beta E_A} \omega \left( E_A-U \left( {\bf r}^n \right) \right)} \nonumber \\
&=
\frac{e^{-\beta \left(E_B-U \left( {\bf r}^n \right)\right)} \left( E_B-U \left( {\bf r}^n \right) \right)^{\sfrac{N_{\text{DOF}}}{2}-1}}
     {e^{-\beta \left(E_A-U \left( {\bf r}^n \right)\right)} \left( E_A-U \left( {\bf r}^n \right) \right)^{\sfrac{N_{\text{DOF}}}{2}-1}} \label{eq:detailedbalance6}
\end{align}
in which we used the known expression for the ideal gas microcanonical partition
function.\cite{tuc101} Eq.~\ref{eq:detailedbalance6} is satisfied by the
\gls{csvr} thermostat, which rescales velocities to the target kinetic
energy distribution given by the gamma distribution:
\begin{equation}
\label{eq:gamma}
  P(K) =\frac{e^{-\beta K} K^{\sfrac{N_{\text{DOF}}}{2}-1}}{\int_0^{\infty}
  dK K^{\sfrac{N_{\text{DOF}}}{2}-1} e^{-\beta K}}
  =\frac{e^{-\beta K} K^{\sfrac{N_{\text{DOF}}}{2}-1}}
  {\beta^{-\sfrac{N_{\text{DOF}}}{2}}\Gamma\left(\sfrac{N_{\text{DOF}}}{2}\right)}
\end{equation}
Hence, the \gls{csvr} thermostat satisfies detailed balance.

\subsubsection{Velocity rescaling to other kinetic energy distributions}

We have seen that simple velocity rescaling violates balance and brings
about the flying ice cube effect, while the \gls{csvr} thermostat
satisfies detailed balance and does not exhibit the artifact.  One key
difference between these algorithms is that simple velocity rescaling
restricts the rescaling factor ($\lambda$) to be less than one when the
system's instantaneous temperature is greater than the target
temperature and greater than one when the instantaneous temperature is
less than the target temperature.  It is this restriction which allowed
us to show graphically that simple velocity rescaling moves decrease
accessible phase space.  It is instructive to consider the effects of
relaxing this restriction while rescaling velocities to a non-canonical
kinetic energy distribution. This procedure would not render any areas
of phase space inaccessible, but the rescaling would be to a
distribution that is not necessarily invariant under Hamiltonian
dynamics.\cite{and801,man991}

\begin{figure}
\centering
\includegraphics[width=3.46in]{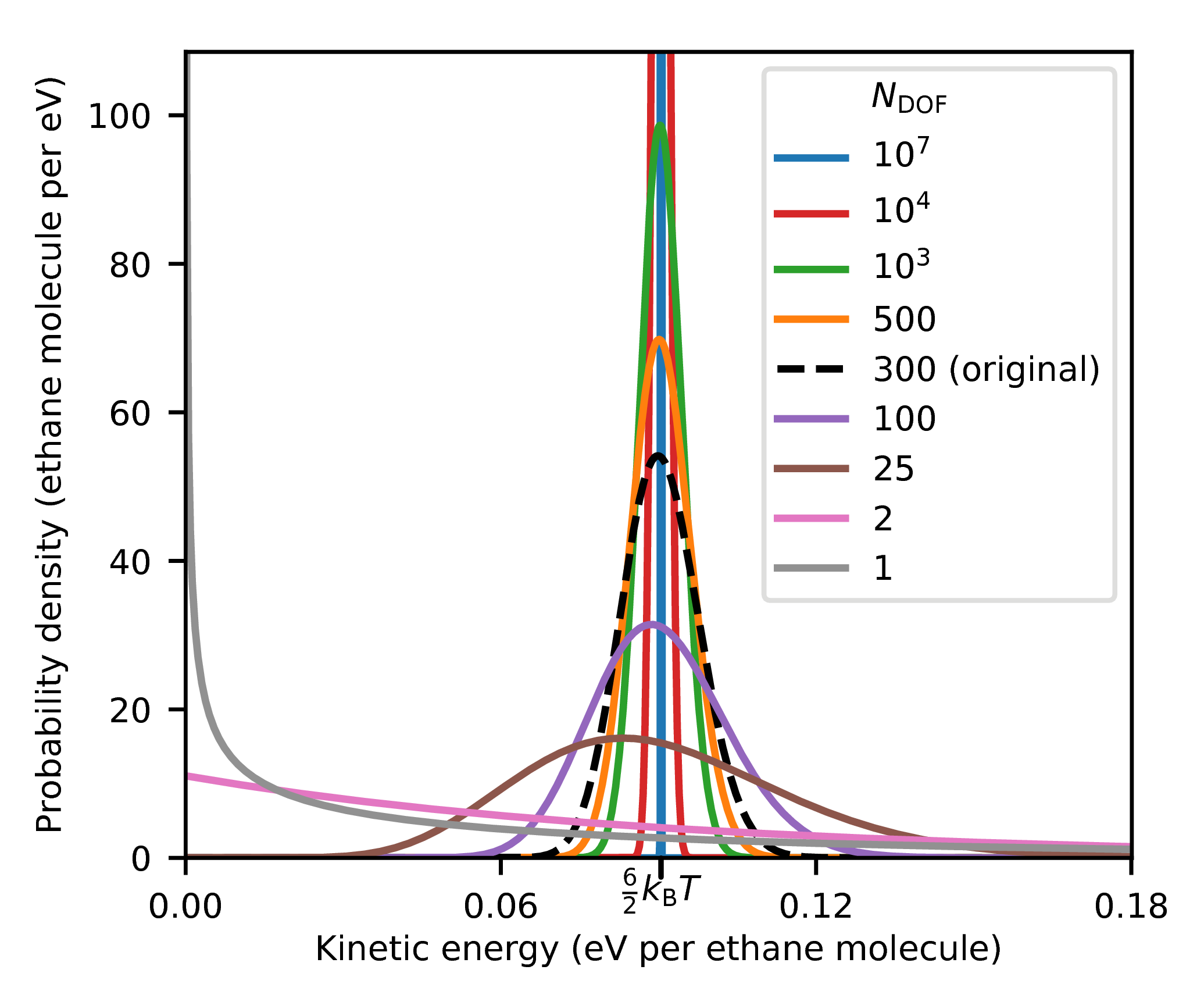}
\\
\begin{minipage}[b]{\textwidth}
\includegraphics[width=3.67in]{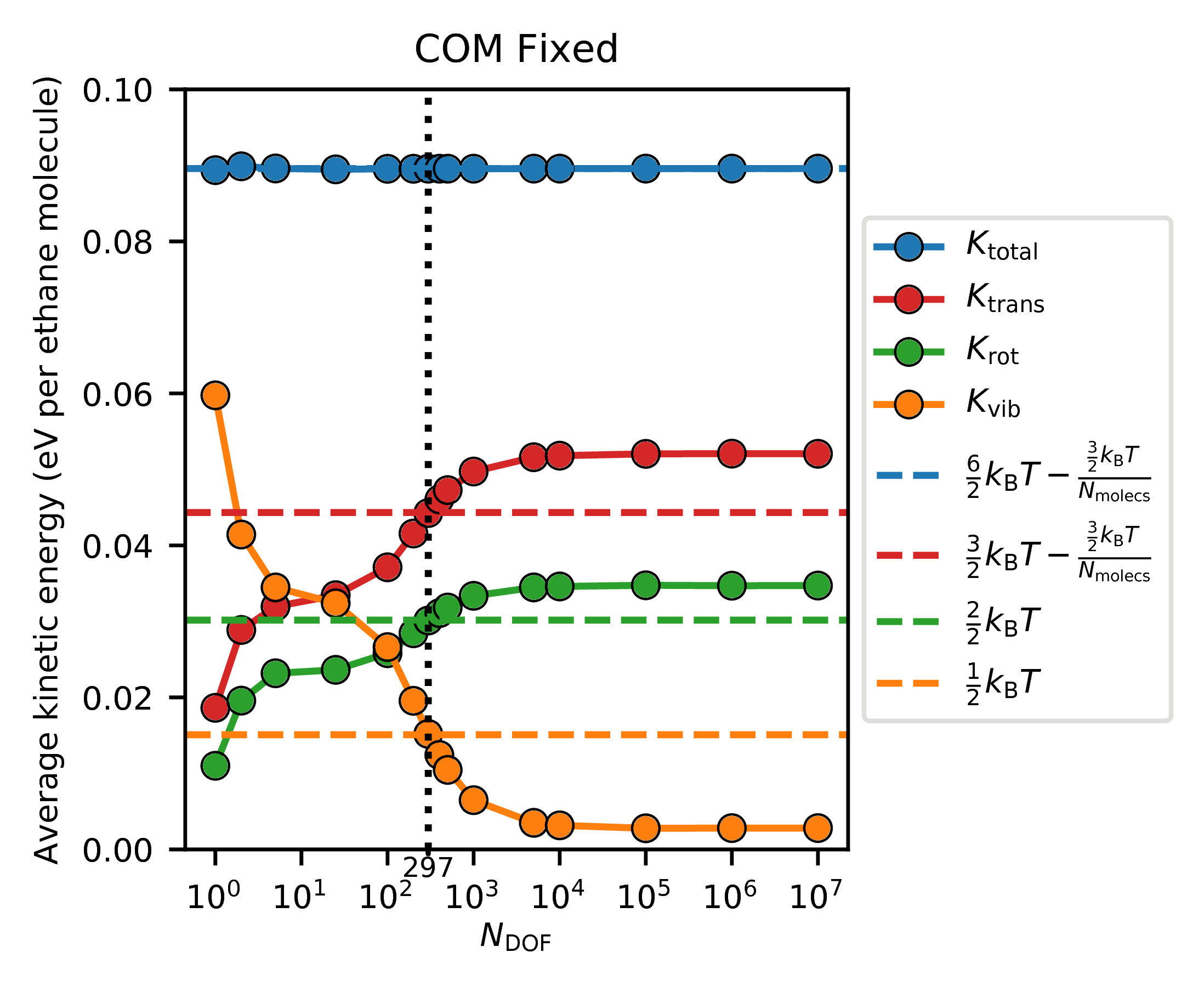}
~
\includegraphics[width=3.33in]{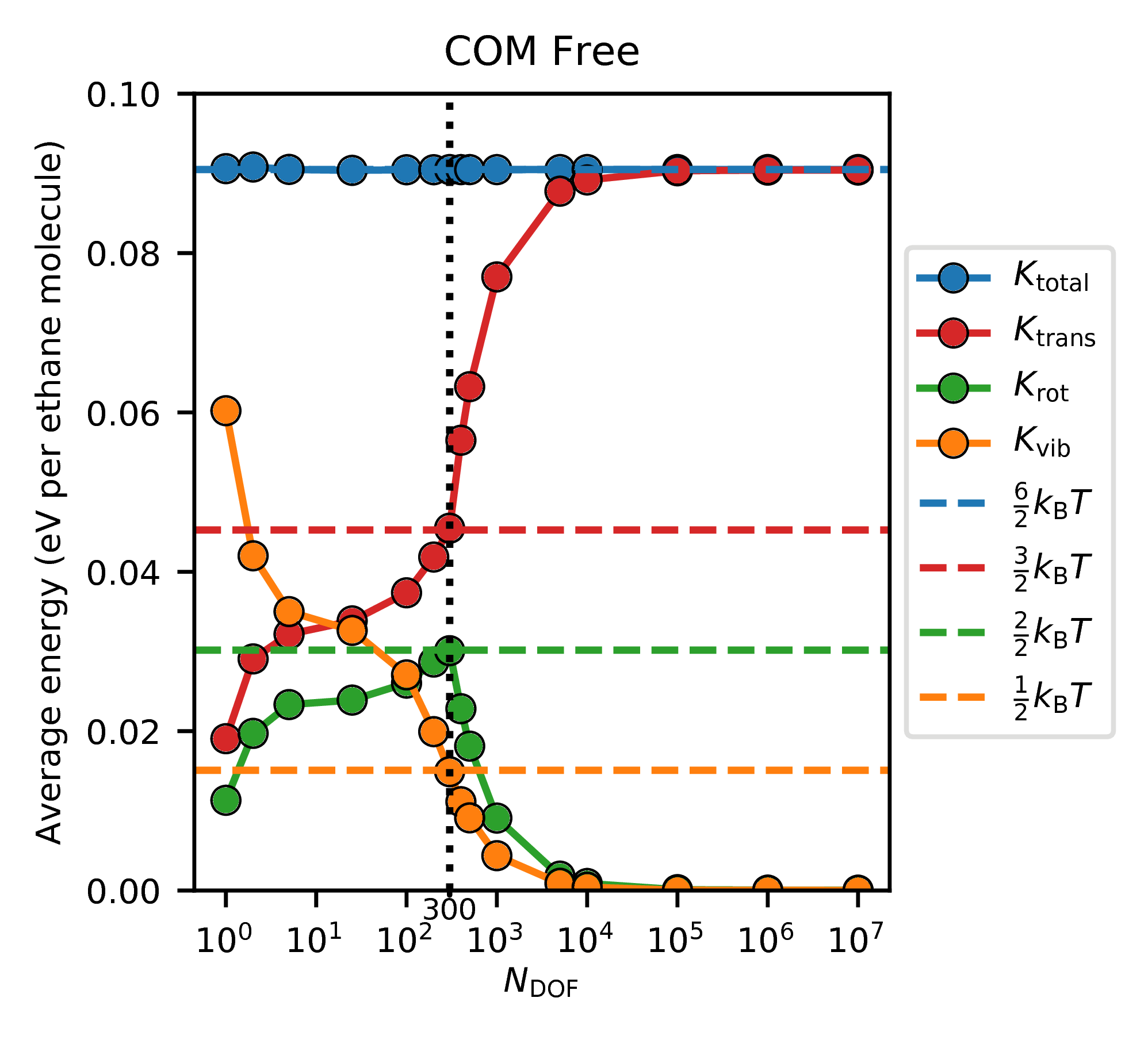}
\end{minipage}
  \caption{\label{fig:bussidof}
    (top) Probability density function of kinetic energies following
    $P(K) =\frac{e^{-\beta K} K^{\sfrac{N_{\text{DOF}}}{2}-1}}
    {\beta^{-\sfrac{N_{\text{DOF}}}{2}}\Gamma\left(\sfrac{N_{\text{DOF}}}{2}\right)}$,
    where $\beta$ is chosen such that the average kinetic energy
    (temperature) is the same for all choices of $N_{\text{DOF}}$ via
    $\beta=\beta_0\frac{N_{\text{DOF}}}{N_{\text{DOF},0}}$,
    $N_{\text{DOF},0}=300$, and
    $\beta_0=\left(k_{\text{B}}\times350\text{ K}\right)^{-1}$.
    (bottom) Partitioning of the kinetic energies obtained from \gls{md}
    simulations of 50 ethane molecules in a \SI{30}{\angstrom} cubic
    simulation box using the \gls{csvr} thermostat, modified such that
    the target distribution of kinetic energies was set to those shown
    in the top part of the figure for the proper $N_{\text{DOF},0}$
    value.
    (bottom left) Here, the \gls{com} momentum was fixed at zero
    and $N_{\text{DOF},0}$ was set to 297.
    (bottom right) Here, the \gls{com} momentum was not fixed after
    the Langevin thermostat equilibration, allowing the \gls{com}
    momentum to drift, and $N_{\text{DOF},0}$ was set to 300.
    Lines are a guide to the eye.}
\end{figure}

To change the target kinetic energy distribution, we modified the
\gls{csvr} thermostat's value of $N_{\text{DOF}}$ in Eq.~\ref{eq:gamma}
from the actual number of \acrlongpl{dof} ($N_{\text{DOF},0}$) while
simultaneously adjusting $\beta$ from its initial value ($\beta_0$) such
that $\beta=\beta_0\frac{N_{\text{DOF}}}{N_{\text{DOF},0}}$ in order to
maintain a constant average kinetic energy. The resulting kinetic energy
distributions are shown in the top of Fig.~\ref{fig:bussidof} and
include distributions that are sharper
($N_{\text{DOF}}>N_{\text{DOF},0}$) and broader
($N_{\text{DOF}}<N_{\text{DOF},0}$) than the canonical distribution.  In
the limit of $N_{\text{DOF}}\to\infty$, this method closely approximates
simple velocity rescaling or the Berendsen thermostat, depending on the
time damping constant used.

The energy partitionings that resulted from setting these target kinetic
energy distributions are shown for simulations in the bottom of
Fig.~\ref{fig:bussidof}.  It can be seen that with sharper
distributions, the flying ice cube effect is observed, with more kinetic
energy partitioned in low-frequency modes and less in high-frequency
modes. Interestingly, the opposite effect is observed with broader
distributions, with more kinetic energy partitioned in high-frequency
modes and less in low-frequency modes.  When the \gls{com} momentum is
not constrained to zero, a more drastic effect is observed, such that
rotational kinetic energy decreases both with decreasing
$N_{\text{DOF}}$ as energy flows to the higher-frequency vibrational
modes and with increasing $N_{\text{DOF}}$ as almost all energy flows to
the lower-frequency translational modes.  Only at the canonical kinetic
energy distribution ($N_{\text{DOF}}=297$ and $N_{\text{DOF}}=300$ for
the constrained and not-constrained \gls{com} momentum simulations,
respectively) is proper equipartitioning observed.

\subsection{Conditions affecting the flying ice cube effect's conspicuousness}

Artifacts relating to the flying ice cube effect do not always appear
when the simple velocity rescaling or Berendsen thermostat algorithms
are used.\cite{mar021,mud041,bas131} Indeed, when the flying ice effect
was first found,\cite{lem941,har981} fewer alternatives to these
thermostatting algorithms were available than at present, e.g., the
\gls{csvr} thermostat had not yet come into popular use, and so protective
measures were recommended to lower the likelihood of the artifact
occurring under these faulty thermostats.\cite{har981} Here, we
investigate these recommendations and other conditions which we found
affect the conspicuousness of the flying ice cube effect for our system
of interacting diatomic ethane molecules.

\begin{figure}
  \centering
  \includegraphics[width=6.87in]{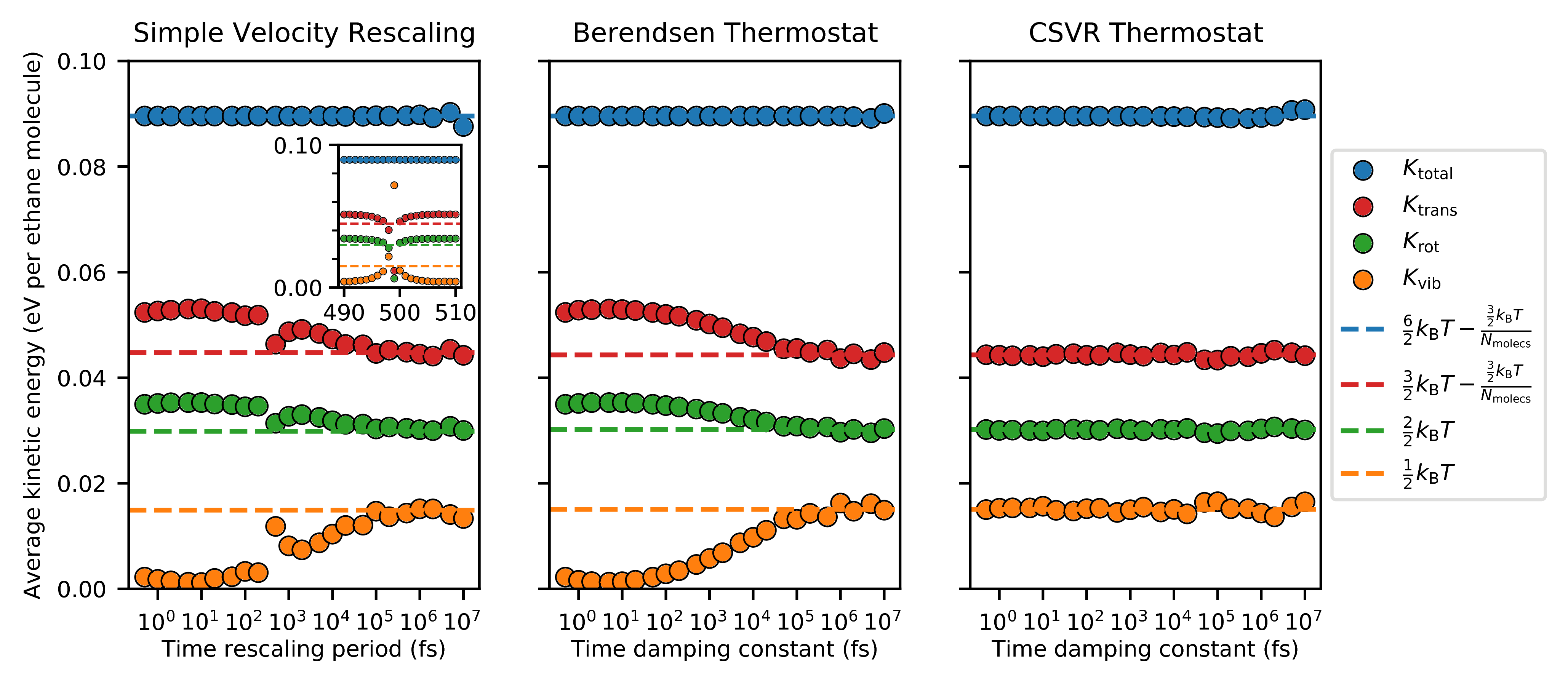}
  \caption{\label{fig:dampingconst}
    Partitioning of the kinetic energies obtained from \gls{md}
    simulations performed under the same conditions as in
    Fig.~\ref{fig:50partsequipartition}, but changing (left) the time
    rescaling period for simple velocity rescaling, (middle) the time
    damping constant for the Berendsen thermostat, and (right) the time
    damping constant for the \gls{csvr} thermostat, all three with the
    timestep maintained at \SI{0.5}{\femto\second}. The inset shown in
    the simple velocity rescaling graph shows additional data near the
    time rescaling period of \SI{500}{\femto\second}, at which point a
    resonance artifact associated with the \ce{CH3-CH3} bond's
    characteristic vibrational frequency can be observed.}
\end{figure}

One recommendation given in \citet{har981} was to lower the thermostat's
coupling strength, either by less frequent rescaling under simple
velocity rescaling or by increasing the time damping constant under the
Berendsen thermostat.  Decreasing the coupling strength allows for the
system's natural dynamics to bring about energy equipartitioning faster
than the thermostat can disturb it.  In Fig.~\ref{fig:dampingconst}, we
show that this recommendation does indeed reduce the violation of
equipartition.  However, the flying ice cube artifact was not fully
resolved until these time parameters were larger than
\SI{100}{\pico\second}, a value much greater than the
\SI{0.5}{\pico\second} time damping constant above which \citet{ber841}
showed that energy fluctuations under the Berendsen thermostat are
similar to energy fluctuations in the microcanonical ensemble and thus
concluded that the thermostat has little influence on the dynamics. This
discrepancy may be partially explained by the use of the rigid SPC water
model\cite{ber811} to evaluate the Berendsen thermostat in
\citet{ber841}, as a rigid molecule lacks the high-frequency vibrational
modes that lead most directly to the flying ice cube effect.  Meanwhile,
we found that energy equipartitioning held under the \gls{csvr}
thermostat regardless of the value of the time damping constant. At the
weakest coupling strengths shown in Fig.~\ref{fig:dampingconst}, it can
be seen that the desired temperature was not well established in these
\SI{100}{\nano\second} simulations.

Varying the coupling strength does not come without its risks.
Fig.~\ref{fig:dampingconst} shows an anomalous data point when simple
velocity rescaling is performed every \SI{500}{\femto\second}.  Further
investigation allowed us to characterize this anomaly as a resonance
effect associated with bond vibration. The characteristic period of the
\ce{CH3-CH3} harmonic bond is \SI{38.4}{\femto\second}. When the time
rescaling period is set close to an integer multiple of half this
characteristic period, large amplitude bond vibrations occur, becoming
stronger when the time rescaling period more exactly matches the
multiple. These resonance effects become weaker as the multiple grows,
which explains why the vibrational energy at the time rescaling period
of \SI{1000}{\femto\second} is greater than at \SI{2000}{\femto\second}.
We observed resonance effects when rescaling close to other multiples of
half the bond's characteristic period that we also tested.
We will shortly show that altering the coupling strength can bring about
resonance effects under the Berendsen thermostat as well.

\begin{figure}
  \centering
  \includegraphics[width=6.53in]{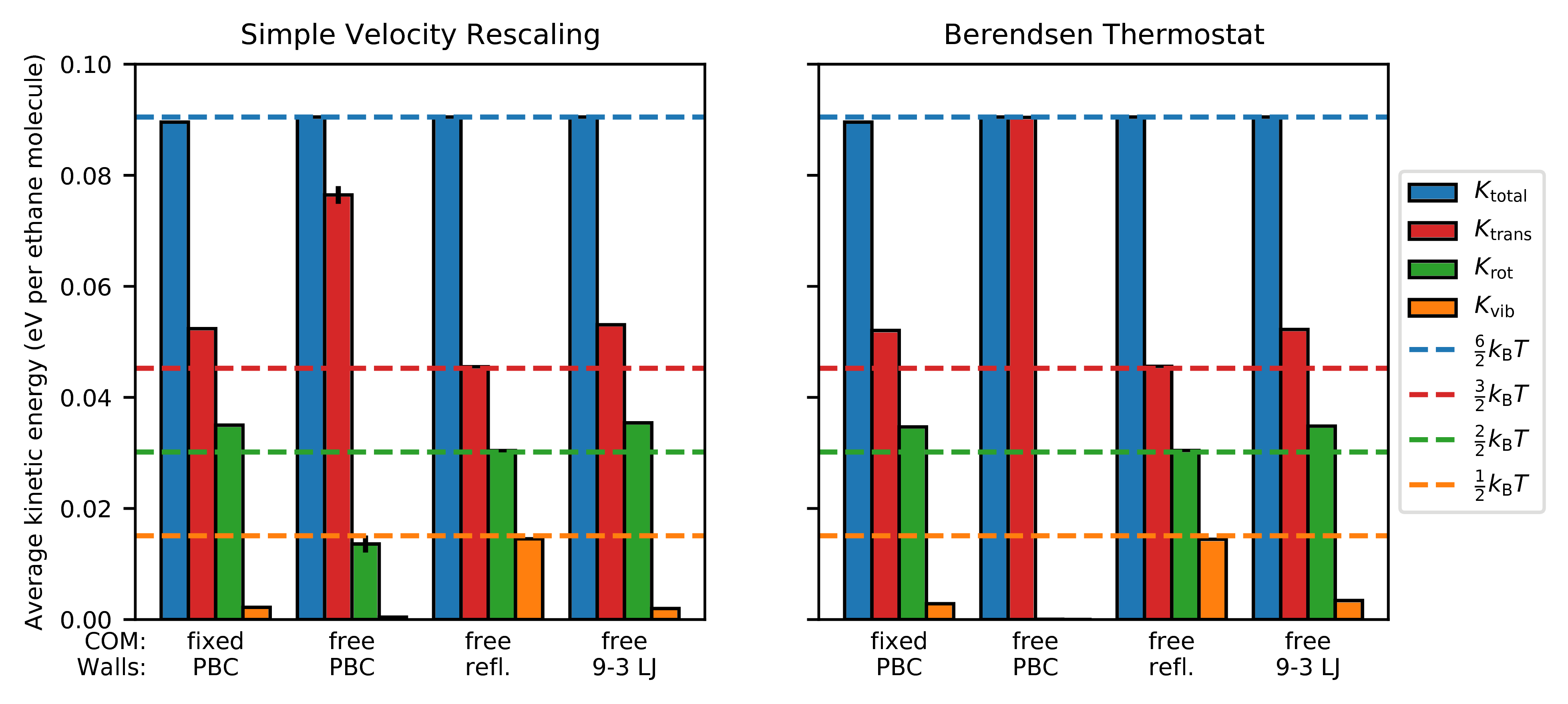}
  \caption{\label{fig:50partsequipartitionberendsen}
    Partitioning of the kinetic energies obtained from \gls{md}
    simulations of 50 ethane molecules in a \SI{30}{\angstrom} cubic
    simulation box under different conditions using (left) simple
    velocity rescaling and (right) the Berendsen thermostat.  In each,
    the first simulation from left is the same simulation as shown in
    Fig.~\ref{fig:50partsequipartition} and provides a basis for
    comparison.  The second simulation shows the effects of letting the
    \gls{com} momentum drift (COM: free) as opposed to fixing it to zero
    (COM: fixed).  The third and fourth simulations show the effects of
    hard (PBC: reflecting) and soft (PBC: 9-3 Lennard-Jones) wall
    boundaries, respectively, as opposed to \gls{pbc} (Walls: PBC).
    Note that the dashed lines meant as a guide to the eye do not
    include the \gls{com} momentum constraint correction of
    $\frac{\frac{3}{2}k_{\text{B}} T}{N_{\rm{molecs}}}$ that is reflected in
    the first simulation.}
\end{figure}

Another precautionary measure recommended in \citet{har981} was to
periodically zero the \gls{com} momentum, as it represents the
lowest-frequency \acrlong{dof} into which most kinetic energy flows.  The
Newtonian equations of motion preserve \gls{com} momentum, but numeric
errors cause this preservation to be inexact. Constraint of the
\gls{com} momentum to zero is oftentimes used to safeguard against these
numeric errors: a safeguard we used throughput this paper except where
stated.  In Fig.~\ref{fig:50partsequipartitionberendsen}, we show that
releasing this constraint does indeed significantly worsen the flying
ice cube effect, though equipartition is violated both with and without
the constraint.  We further explored the effects of allowing the
\gls{com} momentum to vary by replacing the \gls{pbc} with reflecting
walls, which we found gets rid of the flying ice cube effect completely,
with no violation of the equipartition theorem.  In both of these cases,
\gls{com} momentum is not conserved, but with opposite results observed
(though in the former case, \gls{com} momentum can build-up, while in
the latter case, it cannot), We hypothesize that reflecting walls void
the flying ice cube effect because the additional collisions with the
walls give additional opportunities for energy to be transferred between
kinetic modes, which acts more quickly than the Berendsen thermostat
works to incorrectly partition the energy. To test this hypothesis, we
made the walls softer so that a smaller redistribution of intramolecular
kinetic energy would take place upon collision. Instead of reflecting
walls, we used wall-particle interactions with a softer 9-3
Lennard-Jones potential,\cite{ste731}
$U(r)=\epsilon\left[\frac{2}{15}\left(\frac{\sigma}{r}\right)^{9}-\left(\frac{\sigma}{r}\right)^{3}\right]$
with arbitrary $\epsilon$ and $\sigma$ values of
\SI{0.195}{\kcal\per\mole} and \SI{3.75}{\angstrom}, respectively, and a
shifted cutoff of \SI{14}{\angstrom}. We found that with this softer
wall, energy equipartitioning holds less well than with the harder wall,
giving some support to our hypothesis.  We note further that the
presence of the reflecting wall did not significantly change the
distribution of total kinetic energies, i.e., the wall did not bring
about equipartition indirectly through bringing about a more proper
kinetic energy distribution.

\begin{figure}
  \centering
  \includegraphics[width=6.53in]{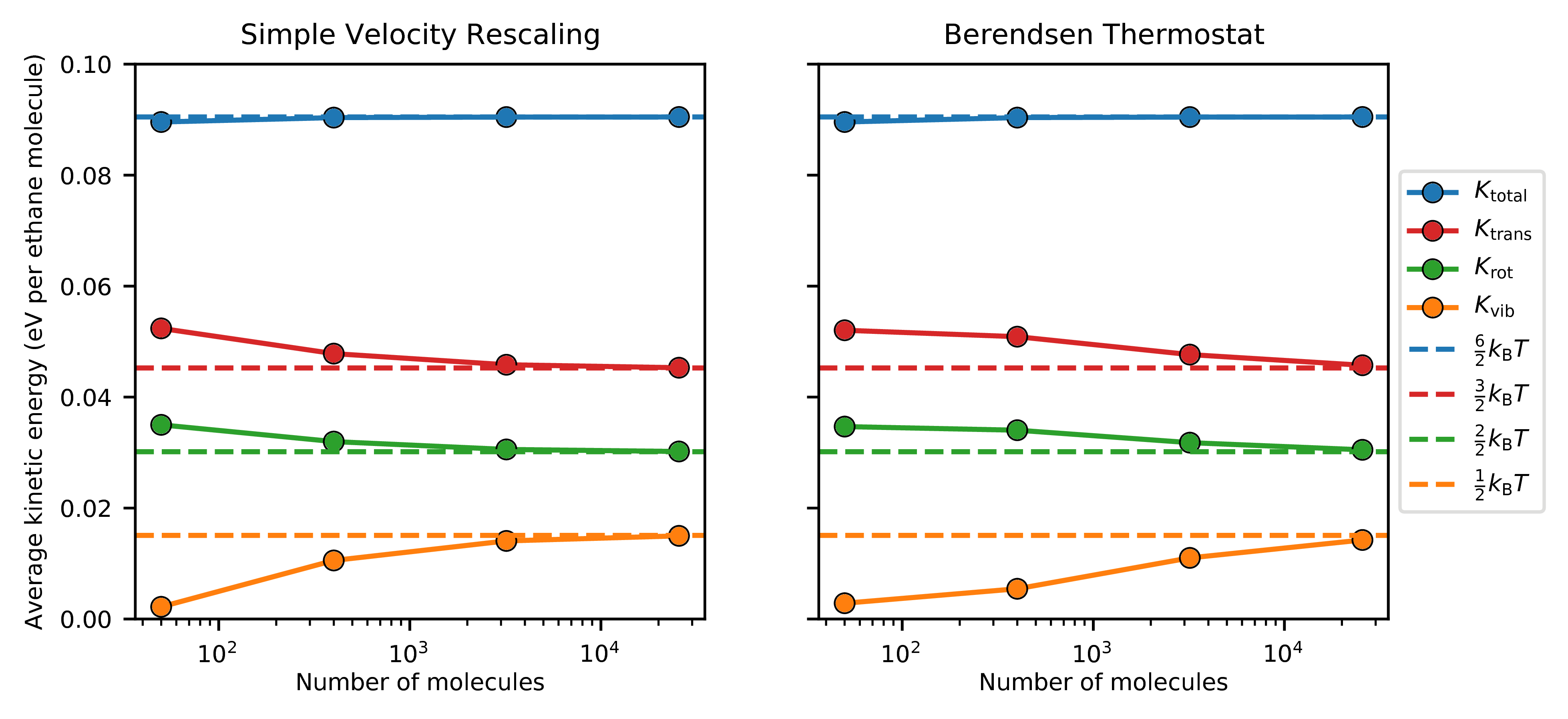}
  \caption{\label{fig:numparts}
    Partitioning of the kinetic energies obtained from \gls{md}
    simulations performed under the same conditions as in
    Fig.~\ref{fig:50partsequipartition} but changing the number of
    ethane molecules, using (left) simple velocity rescaling and (right)
    the Berendsen thermostat. The simulation with 50 ethane molecules
    took place in a \SI{30}{\angstrom} cubic simulation box, and the
    other simulations had their simulation boxes enlarged to maintain
    the same density.  Note that the dashed lines meant as a guide to
    the eye do not include the \gls{com} momentum constraint correction
    of $\frac{\frac{3}{2}k_{\text{B}} T}{N_{\rm{molecs}}}$, which is
    responsible for the slight deviation of the total kinetic energy from
    $\frac{6}{2}k_{\text{B}}T$ that is more evident for the simulations with
    less molecules.}
\end{figure}

Finally, we found that increasing the size of the simulation box reduces
the flying ice cube effect, as can be seen in Fig.~\ref{fig:numparts}.
As with decreasing the timestep (Fig.~\ref{fig:timestep}), here too we
find that simple velocity rescaling recovers equipartition more easily
than the Berendsen thermostat.  We conjecture that this finite size
effect occurs because the canonical ensemble's distribution of kinetic
energy becomes more sharply peaked with increasing number of particles,
i.e., the ratio of the standard deviation to the mean of the canonical
kinetic energy distribution (the gamma distribution given in
Eq.~\ref{eq:gamma}) scales as $\orderof
\left(\frac{1}{\sqrt{N_{\text{DOF}}}}\right)$ at constant temperature.
Thus, as the number of particles increases, simple velocity rescaling
and the Berendsen thermostat become more similar to the \gls{csvr}
thermostat.

\subsection{Sampling configurational \acrlongpl{dof}}

So far, we have exclusively used kinetic \acrlongpl{dof} to show that
the simple velocity rescaling and Berendsen thermostat algorithms cause
the violation of equipartition. These methods are sometimes used only to
sample configurational \acrlongpl{dof}, justified on the grounds that
the isokinetic ensemble samples the same configurational phase space as
the canonical ensemble.\cite{hai831,eva832,nos842,min031,col101} Since
we have proven that the violation of equipartition is incommensurate
with sampling the isokinetic ensemble, it follows that this
justification is invalid.  We now wish to show this explicitly. To do
so, we will examine the \gls{rdf}, which is solely dependent on
configurational \acrlongpl{dof}.

\begin{figure}
  \centering
  \begin{minipage}{0.49\textwidth}
  \includegraphics[width=3.4in]{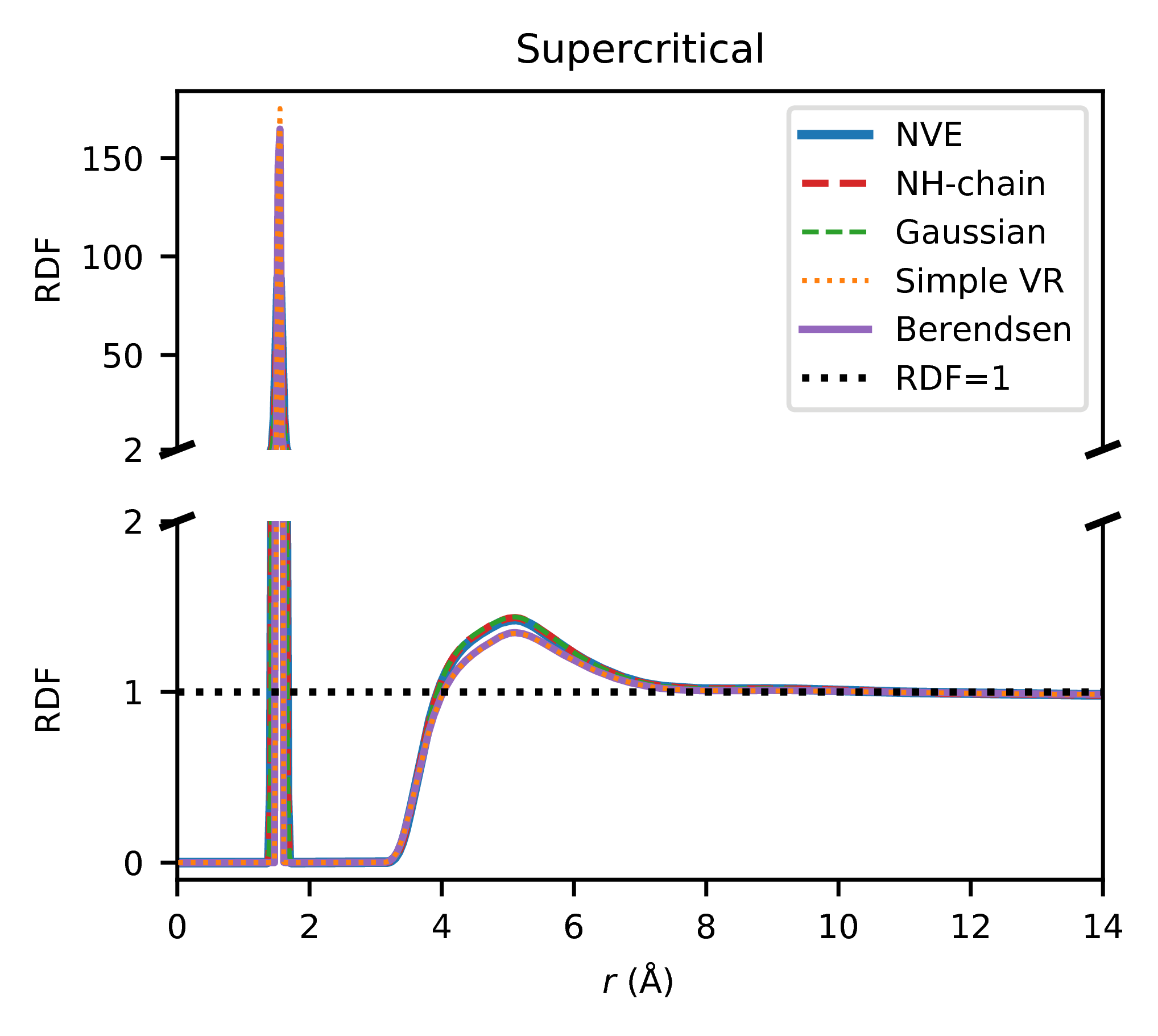}
  \end{minipage}
  \begin{minipage}{0.49\textwidth}
  \includegraphics[width=3.34in]{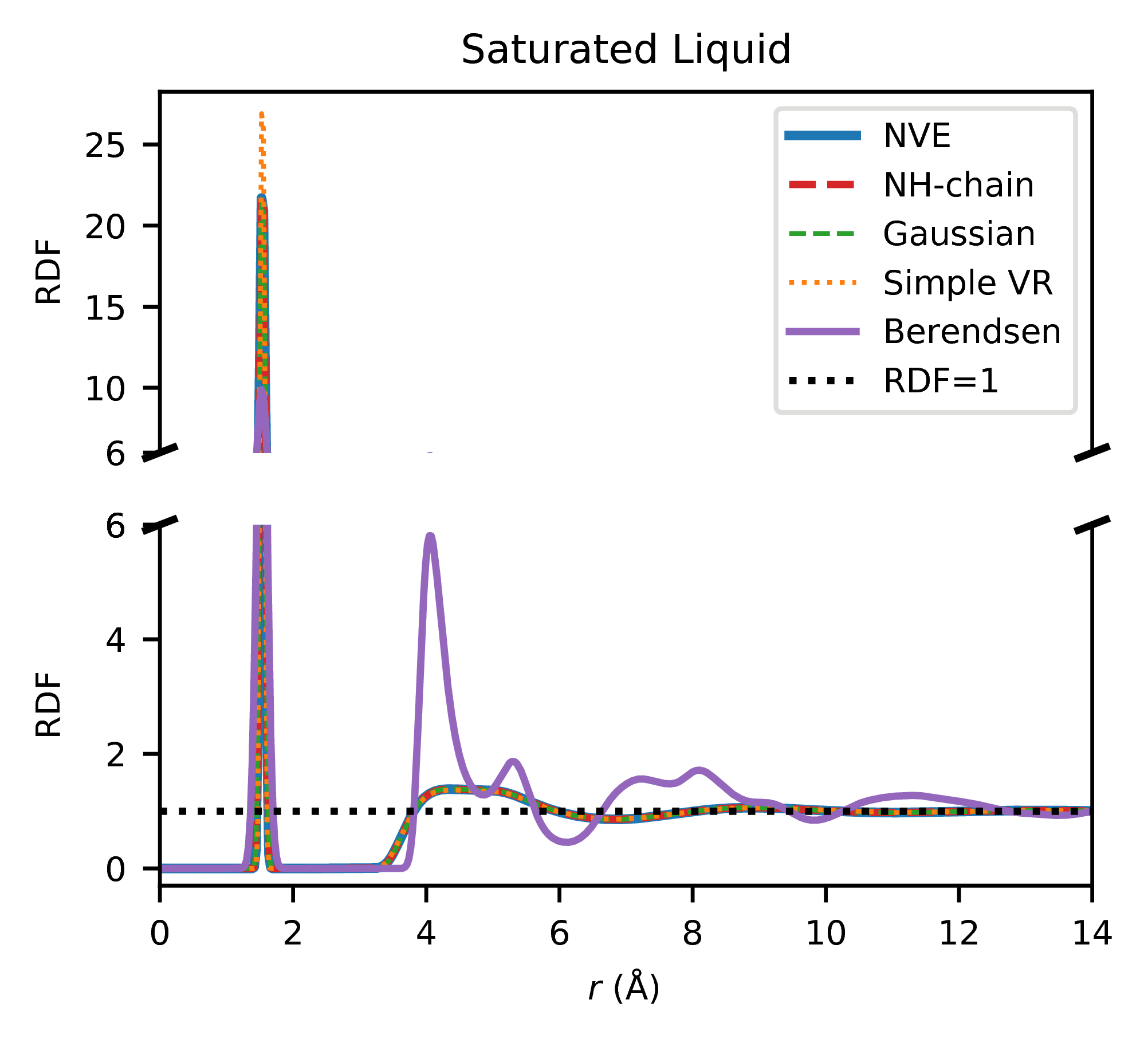}
  \end{minipage}
  \\
  \includegraphics[width=5.14in]{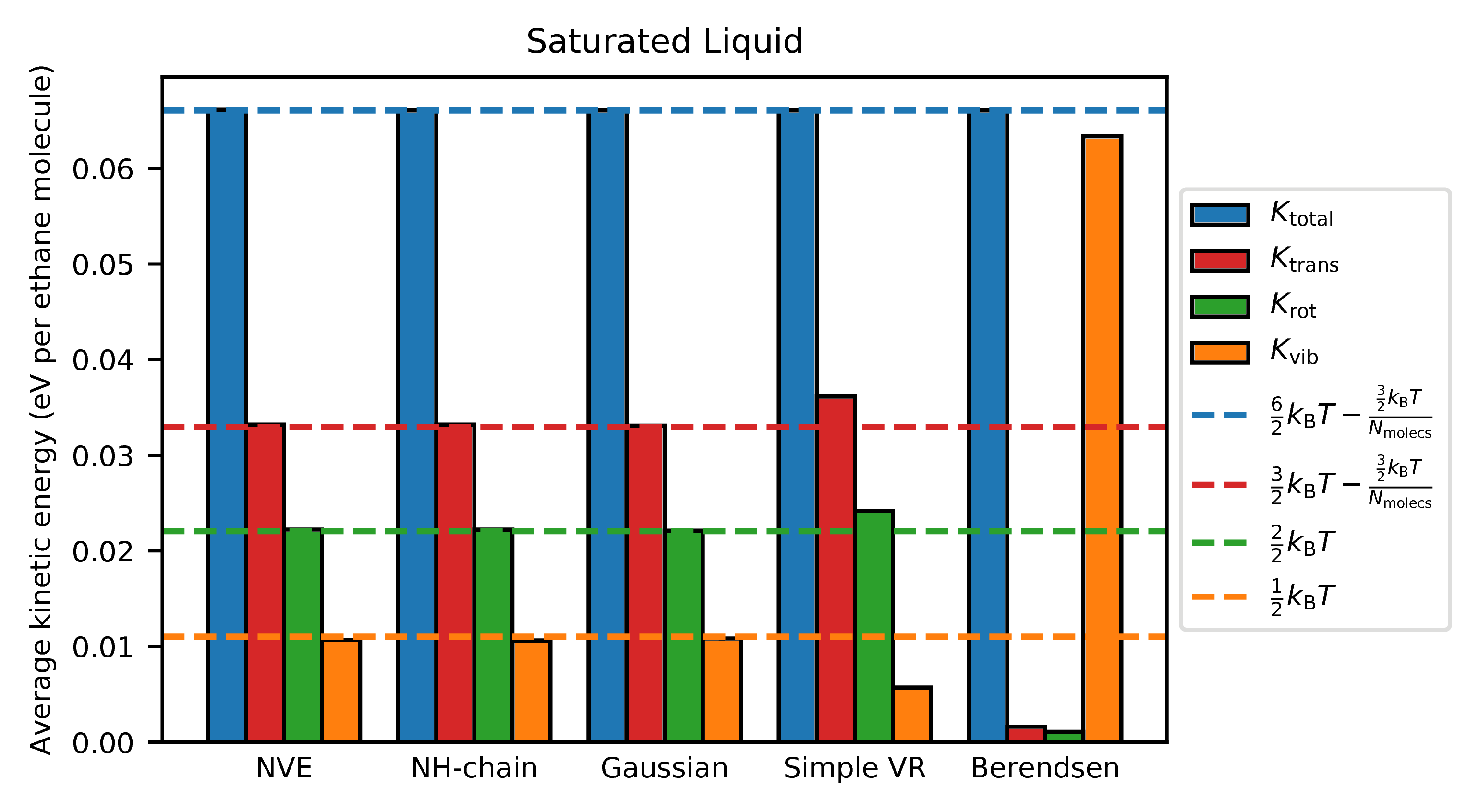}
  \caption{\label{fig:rdf}
    (top-left) \Acrfull{rdf} of the \ce{CH3-CH3}
    distance obtained from the \gls{md} simulations of 50 ethane
    molecules in a \SI{30}{\angstrom} cubic simulation box with a target
    temperature set to \SI{350}{\kelvin} using various thermostats.
    These simulations were the same as the ones whose kinetic energy
    partitionings are shown in Fig.~\ref{fig:50partsequipartition}.
    (top-right) \gls{rdf} of the \ce{CH3-CH3} distance obtained from
    \gls{md} simulations of 235 ethane molecules in a \SI{30}{\angstrom}
    cubic simulation box with a target temperature set to
    \SI{256}{\kelvin} using various thermostats. These conditions were
    chosen such that the simulation would take place under saturated
    liquid conditions.\cite{mar982} For both sets of simulations,
    \gls{com} momentum was fixed to zero throughout.  The \glspl{rdf} of
    both sets of simulations done using the Langevin and \gls{csvr} thermostats
    were indistinguishable from the \gls{rdf} using the \gls{nosehoover}
    thermostat within the line width.
    (bottom) Partitioning of the kinetic energies obtained from the
    saturated liquid simulations. The results of the simulations using the
    Langevin and \gls{csvr} thermostats were indistinguishable from the dashed
    lines of equipartition within the line width.}
\end{figure}

In Fig.~\ref{fig:rdf} (top-left), we show the \glspl{rdf} of the
supercritical ethane simulations whose kinetic energy partitionings are
shown in Fig.~\ref{fig:50partsequipartition}.  The \gls{nosehoover},
\gls{csvr}, Langevin, and Gaussian thermostat simulations exhibit
identical \glspl{rdf}, but the simple velocity rescaling and Berendsen
thermostat simulations show a subtly different \gls{rdf}. Although the
difference is slight, it is sufficient to demonstrably disprove the
claims that simple velocity rescaling samples the same configurational
phase space as the canonical ensemble and that the Berendsen thermostat
samples a configurational phase space intermediate between the canonical
and microcanonical ensembles.\cite{mor001,mor031}

We next turn to saturated liquid phase ethane simulations, for which we
show \glspl{rdf} under various thermostats in Fig.~\ref{fig:rdf}
(top-right).  The \gls{nosehoover}, Langevin, \gls{csvr}, and Gaussian
thermostats all give identical results typical of a simple diatomic
liquid.\cite{cha871} The simple velocity rescaling algorithm once again
shows a subtle difference, but the Berendsen thermostat shows a very
different \gls{rdf} more reminiscent of the solid phase than the liquid
phase,\cite{cha871} and visualization of the Berendsen thermostat system
shows that the ethane molecules have indeed packed into a volume smaller
than available in the simulation box. Examination of the kinetic energy
partitionings in Fig.~\ref{fig:rdf} (bottom) shows that most of the
kinetic energy is in vibrational modes, which is unexpected since that
is the opposite of the usual flying ice cube result. The Berendsen
thermostat's results are heavily dependent on the choice of time damping
constant, with
the \gls{rdf} indicating a solid-like phase for time damping constants
approximately from \SIrange{10}{150}{\femto\second}
(Fig.~\ref{fig:rdfSI}).
This effect of intermediate time damping constants giving larger
deviations than small or large ones has been observed before in simulations
of bulk water, where the effect was attributed to the intermediate time
constant matching a characteristic time scale on which dynamical
correlations are most pronounced.\cite{mud041}
It appears clear that the Berendsen thermostat is not immune to
the resonance artifacts that we have also seen with simple velocity
rescaling (Fig.~\ref{fig:dampingconst}).

\subsection{Contemporary use of the simple velocity rescaling and
Berendsen thermostat algorithms}

Ours is not the first publication to warn against the use of simple
velocity rescaling and the Berendsen
thermostat.\cite{har981,coo081,shi131} Nonetheless, as we have stated,
these algorithms continue to be widely used
(Fig.~\ref{fig:thermostatcitations}).  As we have just shown, for some
systems the improper velocity rescaling algorithms may not greatly
affect the system properties, and there are a slew of studies in which
these thermostats are tested for specific systems, with some showing
artifacts and others showing
indistinguishability.\cite{che961,mar021,mud041,mor081,ros091,spi111,bas131}
However, slight changes to a system could introduce artifacts in an
unpredictable fashion.
Rather than testing for the correctness of simple velocity rescaling or
the Berendsen thermostat in every specific system, we advocate for the
cessation of their use.  We find no reason to use simple velocity
rescaling or the Berendsen thermostat instead of the \gls{csvr}
thermostat given their similar ease of implementation, likely similar
speeds of equilibration,\cite{bus081} and our study's finding that the
\gls{csvr} thermostat does not lead to the flying ice cube effect,
As a case study on the dangers of continuing to use these thermostat
algorithms, we examine a highly-cited study in depth, the replication of
which initially led us to examine the flying ice cube phenomenon.

In 2007, a flexible force field intended for use with \acrshort{mof}-5 was
parameterized,\cite{taf071} and it was shortly thereafter used to study
the confined transport of guest molecules within the
framework.\cite{ami071} The authors were able to replicate the
experimental diffusion coefficient of confined benzene, but they found
that this replicability only held when the \gls{mof} was allowed to be
flexible; when the \gls{mof} atoms were held rigid, the benzene
diffusion coefficient increased by an order of magnitude.
The conclusions of this manuscript are often evoked to question the
validity of the rigid framework assumption that is commonly used in many
\gls{mof} molecular simulation studies.

The finding continues to be accepted since it is known that the effect
of framework flexibility on guest diffusion is complex,\cite{smi081}
though surprise has been expressed\cite{see091} since a rigid lattice
more typically leads to a decrease in the diffusion coefficient for tight
fitting molecules.\cite{smi081} In addition, using a different flexible
force field for \acrshort{mof}-5,\cite{dub071} it was found that flexibility
had little effect on the diffusion coefficient, increasing it by less
than a factor of 1.5.\cite{for091}

\begin{figure}
  \centering
  \includegraphics[width=3.48in]{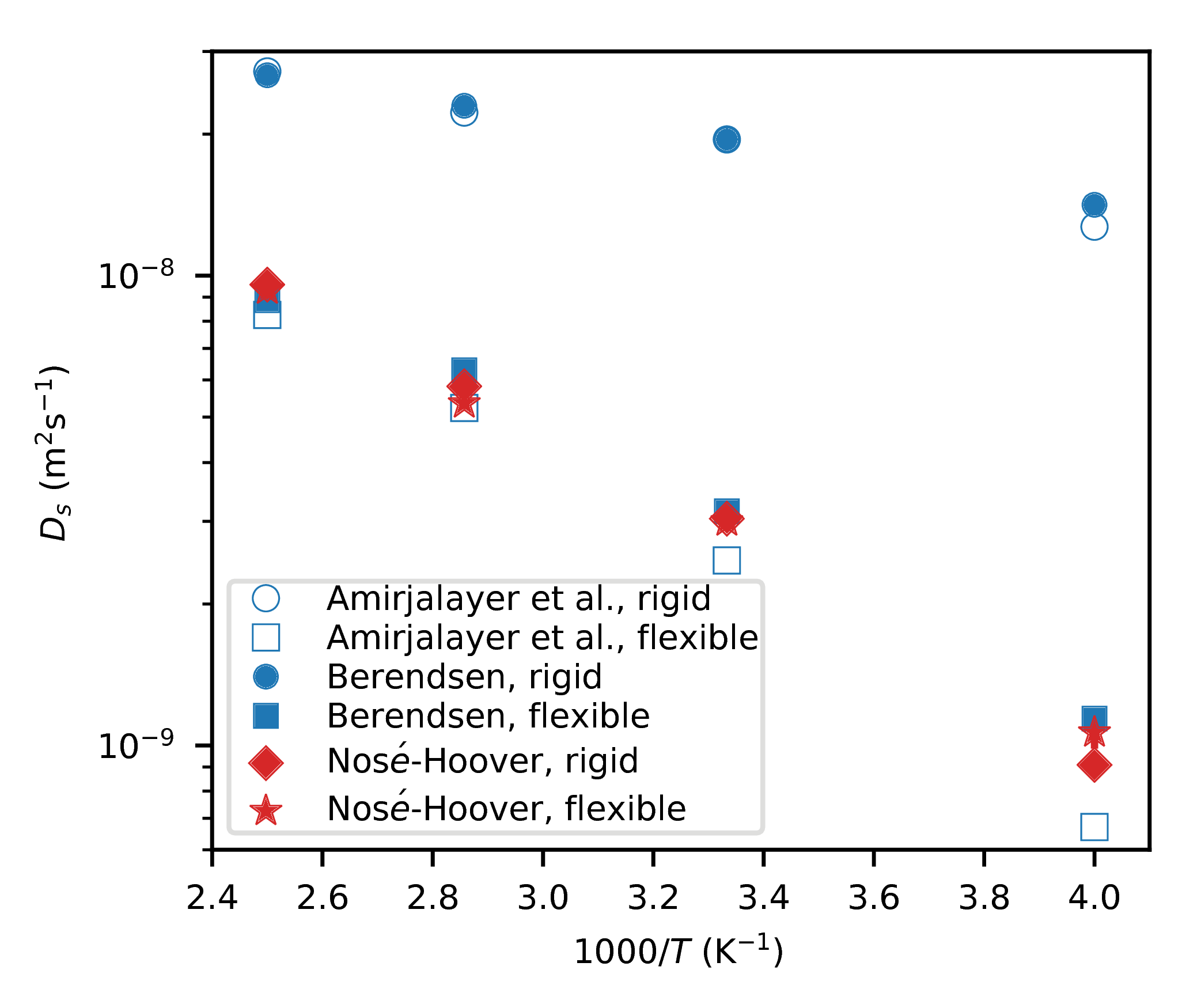}
  \caption{\label{fig:diffcoeff}
    Self-diffusion coefficient of benzene in \acrshort{mof}-5 at a loading of
    10 molecules per unit cell as a function of inverse temperature.
    Data are shown for flexible and rigid frameworks, and using the
    Berendsen and \gls{nosehoover} chain thermostats (use of the \gls{csvr}
    thermostat gives diffusion coefficients that are statistically
    indistinguishable from use of the \gls{nosehoover} thermostat). With
    the Berendsen thermostat, it appears that the framework flexibility
    has a large effect on the calculated diffusion coefficient,
    replicating the main finding of \citet{ami071}.  However, it is seen
    that this result is a flying ice cube artifact, as no flexibility
    effect is seen with the \gls{nosehoover} thermostat.  Error bars
    represent $\pm1$ standard error of the mean using block
    averaging,\cite{fre011} and are not shown for the data from
    \citet{ami071} or if they would be smaller than the symbol size.}
\end{figure}

As the reader now anticipates, \citet{ami071} used the Berendsen
thermostat, which was the default option in the Tinker simulation
package at the time (the default has since been changed to the \gls{csvr}
thermostat).\cite{pon871} As we show in Fig.~\ref{fig:diffcoeff}, the result of
\citet{ami071} was completely an artifact of the Berendsen thermostat.
Using the same force field, no dependence of the benzene diffusion
coefficient on the framework flexibility is observed when a
\gls{nosehoover} or \gls{csvr} thermostat is used.  Apparently, when the
Berendsen thermostat is thermostatted to fewer \acrlongpl{dof} during
rigid framework simulations, the flying ice cube effect becomes more
noticeable and kinetic energy is drawn into the translational modes of
the guest benzene molecules, accounting for the result observed by
\citet{ami071}. We also found that changing the time damping constant of
the Berendsen thermostat had a large effect on the diffusion coefficient
(Fig.~\ref{fig:diffcoeffSI}).

As an aside, it is now known that bulk-like vapor and liquid phases of
benzene exist in \acrshort{mof}-5 below a critical temperature.\cite{bra152}
It is actually improper to calculate the diffusion coefficient at a
loading that is within the vapor-liquid phase envelope, e.g.,
\numrange{3}{67} molecules per unit cell at \SI{300}{\kelvin} in this
system,\cite{bra152} since there is not a single homogeneous phase
present at these conditions.  Here, we are not attempting to calculate
correct diffusion coefficients of benzene in \acrshort{mof}-5, but rather to
compare results with the prior work of \citet{ami071}, which conducted
the simulations at a loading of 10 molecules per unit cell.  The
importance of framework flexibility on the simulated diffusion
coefficient is expected to be independent of the choice of loading.

Other errors, varying in severity, are likely present in many of the
thousands of studies that have used simple velocity rescaling or the
Berendsen thermostat.  Occasionally, one of these errors is explicitly
pointed out,\cite{ley081,won101} but negative replications are not
commonly published,\cite{bak161} so the extent to which these articles
contain data contaminated by the flying ice cube artifact cannot be
estimated.

\section{Concluding Remarks}

In this work, we have shown that rescaling velocities to a non-canonical
distribution of kinetic energies, as is done with the simple velocity
rescaling and Berendsen thermostat algorithms, causes the flying ice
cube effect whereby the equipartition theorem is violated.  Thus, simple
velocity rescaling does not sample the isokinetic ensemble, and the
Berendsen thermostat does not sample a configurational phase space
intermediate between the canonical and microcanonical ensembles;
justifications for their use do not hold.  The flying ice cube effect is
brought about by a violation of balance causing systematic
redistributions of kinetic energy; this violation is lessened as the
timestep between simple velocity rescalings is decreased, eventually
making simple velocity rescaling equivalent to the Gaussian thermostat.
Equipartition violation is completely avoided when velocities are
rescaled to the canonical distribution of kinetic energies, as is done
under the \gls{csvr} thermostat, because detailed balance is obeyed.

We have identified several simulation parameters which affect the
prominence of the flying ice cube effect under simple velocity rescaling
and the Berendsen thermostat. These include the timestep, the
thermostat's coupling strength, the frequency of collisions within the
simulation (e.g., with a wall), and the system size.  However, most of
these parameters cannot be adjusted in a manner that eliminates the
flying ice cube effect without making simulations prohibitively
expensive for relevant systems of contemporary interest.  Another reason
not to attempt to tune these simulation parameters to allow the use of
incorrect thermostatting algorithms is the existence of additional
resonance artifacts that occur when the thermostat coupling strengths
are set to particular values that are difficult to predict \latin{a
priori}.

Finally, we have demonstrated several severe simulation artifacts that
the flying ice cube effect can bring about to the system's structural
and dynamic properties. These include incorrect \glspl{rdf}, phase
properties, and diffusion coefficients. We have highlighted one case in
which the flying ice cube effect has been wholly responsible for the
main finding of a highly-cited study. Many more such cases are likely
present in the literature.

We strongly advocate for discontinuing use of the simple velocity
rescaling and Berendsen thermostat algorithms in all \gls{md}
simulations for both equilibration and production cycles. The results
of past studies that have used these two algorithms should be treated
with caution unless they are shown to be replicable with a more reliable
thermostat. In situations where velocity rescaling methods are
desirable, such as for fast equilibration of a system,\cite{hu041} the
\gls{csvr} thermostat should be used instead.

\begin{acknowledgement}

This research was supported as part of the Center for Gas Separations
Relevant to Clean Energy Technologies, an Energy Frontier Research
Center funded by the U.S. Department of Energy, Office of Science, Basic
Energy Sciences under Award DE-SC0001015. M.M.\ was supported by the
Deutsche Forschungsgemeinschaft (DFG, priority program SPP 1570).  This
research used resources of the National Energy Research Scientific
Computing Center, a DOE Office of Science User Facility supported by the
Office of Science of the U.S. Department of Energy under Contract No.
DE-AC02-05CH11231.  E.B.\ thanks the responders on the LAMMPS mailing
list for useful discussion and for giving advice regarding the LAMMPS
source code (Axel Kohlmeyer, Steven J.\ Plimpton, and Aidan P.\ Thompson
were particularly helpful) and Sai Sanigepalli for helping to implement
the Tinker simulations. Special thanks go to Rochus Schmid for
insightful discussion on the roles of thermostatting and for providing
assistance in implementing the Tinker simulations.

\end{acknowledgement}

\bibliography{flying-ice-cube-arxiv.bbl}

\providecommand{\latin}[1]{#1}
\providecommand*\mcitethebibliography{\thebibliography}
\csname @ifundefined\endcsname{endmcitethebibliography}
  {\let\endmcitethebibliography\endthebibliography}{}
\begin{mcitethebibliography}{72}
\providecommand*\natexlab[1]{#1}
\providecommand*\mciteSetBstSublistMode[1]{}
\providecommand*\mciteSetBstMaxWidthForm[2]{}
\providecommand*\mciteBstWouldAddEndPuncttrue
  {\def\EndOfBibitem{\unskip.}}
\providecommand*\mciteBstWouldAddEndPunctfalse
  {\let\EndOfBibitem\relax}
\providecommand*\mciteSetBstMidEndSepPunct[3]{}
\providecommand*\mciteSetBstSublistLabelBeginEnd[3]{}
\providecommand*\EndOfBibitem{}
\mciteSetBstSublistMode{f}
\mciteSetBstMaxWidthForm{subitem}{(\alph{mcitesubitemcount})}
\mciteSetBstSublistLabelBeginEnd
  {\mcitemaxwidthsubitemform\space}
  {\relax}
  {\relax}

\bibitem[Frenkel and Smit(2002)Frenkel, and Smit]{fre011}
Frenkel,~D.; Smit,~B. \emph{Understanding Molecular Simulation: From Algorithms
  to Applications}; Elsevier Science, 2002\relax
\mciteBstWouldAddEndPuncttrue
\mciteSetBstMidEndSepPunct{\mcitedefaultmidpunct}
{\mcitedefaultendpunct}{\mcitedefaultseppunct}\relax
\EndOfBibitem
\bibitem[Leimkuhler and Matthews(2015)Leimkuhler, and Matthews]{lei151}
Leimkuhler,~B.; Matthews,~C. In \emph{Molecular Dynamics With Deterministic and
  Stochastic Numerical Methods}; Antman,~S.~S., Holmes,~P., Greengard,~L.,
  Eds.; Interdisciplinary Applied Mathematics 39; Springer, 2015\relax
\mciteBstWouldAddEndPuncttrue
\mciteSetBstMidEndSepPunct{\mcitedefaultmidpunct}
{\mcitedefaultendpunct}{\mcitedefaultseppunct}\relax
\EndOfBibitem
\bibitem[Woodcock(1971)]{woo711}
Woodcock,~L.~V. \emph{Chem. Phys. Lett.} \textbf{1971}, \emph{10},
  257--261\relax
\mciteBstWouldAddEndPuncttrue
\mciteSetBstMidEndSepPunct{\mcitedefaultmidpunct}
{\mcitedefaultendpunct}{\mcitedefaultseppunct}\relax
\EndOfBibitem
\bibitem[Evans \latin{et~al.}(1983)Evans, Hoover, Failor, Moran, and
  Ladd]{eva831}
Evans,~D.~J.; Hoover,~W.~G.; Failor,~B.~H.; Moran,~B.; Ladd,~A. J.~C.
  \emph{Phys. Rev. A} \textbf{1983}, \emph{28}, 1016--1021\relax
\mciteBstWouldAddEndPuncttrue
\mciteSetBstMidEndSepPunct{\mcitedefaultmidpunct}
{\mcitedefaultendpunct}{\mcitedefaultseppunct}\relax
\EndOfBibitem
\bibitem[Nos{\'e}(1984)]{nos842}
Nos{\'e},~S. \emph{J. Chem. Phys.} \textbf{1984}, \emph{81}, 511--519\relax
\mciteBstWouldAddEndPuncttrue
\mciteSetBstMidEndSepPunct{\mcitedefaultmidpunct}
{\mcitedefaultendpunct}{\mcitedefaultseppunct}\relax
\EndOfBibitem
\bibitem[Evans and Morriss(1990)Evans, and Morriss]{eva902}
Evans,~D.~J.; Morriss,~G.~P. \emph{Statistical Mechanics of Nonequilibrium
  Liquids}; Theoretical Chemistry Monograph Series; Academic Press: London,
  1990\relax
\mciteBstWouldAddEndPuncttrue
\mciteSetBstMidEndSepPunct{\mcitedefaultmidpunct}
{\mcitedefaultendpunct}{\mcitedefaultseppunct}\relax
\EndOfBibitem
\bibitem[Schneider and Stoll(1978)Schneider, and Stoll]{sch781}
Schneider,~T.; Stoll,~E. \emph{Phys. Rev. B} \textbf{1978}, \emph{17},
  1302--1322\relax
\mciteBstWouldAddEndPuncttrue
\mciteSetBstMidEndSepPunct{\mcitedefaultmidpunct}
{\mcitedefaultendpunct}{\mcitedefaultseppunct}\relax
\EndOfBibitem
\bibitem[Berendsen \latin{et~al.}(1984)Berendsen, Postma, van Gunsteren,
  DiNola, and Haak]{ber841}
Berendsen,~H. J.~C.; Postma,~J. P.~M.; van Gunsteren,~W.~F.; DiNola,~A.;
  Haak,~J.~R. \emph{J. Chem. Phys.} \textbf{1984}, \emph{81}, 3684--3690\relax
\mciteBstWouldAddEndPuncttrue
\mciteSetBstMidEndSepPunct{\mcitedefaultmidpunct}
{\mcitedefaultendpunct}{\mcitedefaultseppunct}\relax
\EndOfBibitem
\bibitem[Bussi \latin{et~al.}(2007)Bussi, Donadio, and Parrinello]{bus071}
Bussi,~G.; Donadio,~D.; Parrinello,~M. \emph{J. Chem. Phys.} \textbf{2007},
  \emph{126}, 014101\relax
\mciteBstWouldAddEndPuncttrue
\mciteSetBstMidEndSepPunct{\mcitedefaultmidpunct}
{\mcitedefaultendpunct}{\mcitedefaultseppunct}\relax
\EndOfBibitem
\bibitem[Heyes(1983)]{hey831}
Heyes,~D.~M. \emph{Chem. Phys.} \textbf{1983}, \emph{82}, 285--301\relax
\mciteBstWouldAddEndPuncttrue
\mciteSetBstMidEndSepPunct{\mcitedefaultmidpunct}
{\mcitedefaultendpunct}{\mcitedefaultseppunct}\relax
\EndOfBibitem
\bibitem[Nos{\'e}(1984)]{nos841}
Nos{\'e},~S. \emph{Mol. Phys.} \textbf{1984}, \emph{52}, 255--268\relax
\mciteBstWouldAddEndPuncttrue
\mciteSetBstMidEndSepPunct{\mcitedefaultmidpunct}
{\mcitedefaultendpunct}{\mcitedefaultseppunct}\relax
\EndOfBibitem
\bibitem[Hoover(1985)]{hoo851}
Hoover,~W.~G. \emph{Phys. Rev. A} \textbf{1985}, \emph{31}, 1695--1697\relax
\mciteBstWouldAddEndPuncttrue
\mciteSetBstMidEndSepPunct{\mcitedefaultmidpunct}
{\mcitedefaultendpunct}{\mcitedefaultseppunct}\relax
\EndOfBibitem
\bibitem[Martyna \latin{et~al.}(1992)Martyna, Klein, and Tuckerman]{mar921}
Martyna,~G.~J.; Klein,~M.~L.; Tuckerman,~M. \emph{J. Chem. Phys.}
  \textbf{1992}, \emph{97}, 2635--2643\relax
\mciteBstWouldAddEndPuncttrue
\mciteSetBstMidEndSepPunct{\mcitedefaultmidpunct}
{\mcitedefaultendpunct}{\mcitedefaultseppunct}\relax
\EndOfBibitem
\bibitem[Andersen(1980)]{and801}
Andersen,~H.~C. \emph{J. Chem. Phys.} \textbf{1980}, \emph{72},
  2384--2393\relax
\mciteBstWouldAddEndPuncttrue
\mciteSetBstMidEndSepPunct{\mcitedefaultmidpunct}
{\mcitedefaultendpunct}{\mcitedefaultseppunct}\relax
\EndOfBibitem
\bibitem[Tobias \latin{et~al.}(1993)Tobias, Martyna, and Klein]{tob931}
Tobias,~D.~J.; Martyna,~G.~J.; Klein,~M.~L. \emph{J. Phys. Chem.}
  \textbf{1993}, \emph{97}, 12959--12966\relax
\mciteBstWouldAddEndPuncttrue
\mciteSetBstMidEndSepPunct{\mcitedefaultmidpunct}
{\mcitedefaultendpunct}{\mcitedefaultseppunct}\relax
\EndOfBibitem
\bibitem[Morriss and Dettmann(1998)Morriss, and Dettmann]{mor981}
Morriss,~G.~P.; Dettmann,~C.~P. \emph{Chaos} \textbf{1998}, \emph{8},
  321--336\relax
\mciteBstWouldAddEndPuncttrue
\mciteSetBstMidEndSepPunct{\mcitedefaultmidpunct}
{\mcitedefaultendpunct}{\mcitedefaultseppunct}\relax
\EndOfBibitem
\bibitem[H{\"u}nenberger(2005)]{hun051}
H{\"u}nenberger,~P.~H. In \emph{Advanced Computer Simulation}; Holm,~C.,
  Kremer,~K., Eds.; Advances in Polymer Science; Springer, 2005; Vol. 173; pp
  104--149\relax
\mciteBstWouldAddEndPuncttrue
\mciteSetBstMidEndSepPunct{\mcitedefaultmidpunct}
{\mcitedefaultendpunct}{\mcitedefaultseppunct}\relax
\EndOfBibitem
\bibitem[Tuckerman(2010)]{tuc101}
Tuckerman,~M. \emph{Statistical Mechanics: Theory and Molecular Simulation};
  Oxford Graduate Texts; Oxford University Press: Oxford, 2010\relax
\mciteBstWouldAddEndPuncttrue
\mciteSetBstMidEndSepPunct{\mcitedefaultmidpunct}
{\mcitedefaultendpunct}{\mcitedefaultseppunct}\relax
\EndOfBibitem
\bibitem[Haile and Gupta(1983)Haile, and Gupta]{hai831}
Haile,~J.~M.; Gupta,~S. \emph{J. Chem. Phys.} \textbf{1983}, \emph{79},
  3067--3076\relax
\mciteBstWouldAddEndPuncttrue
\mciteSetBstMidEndSepPunct{\mcitedefaultmidpunct}
{\mcitedefaultendpunct}{\mcitedefaultseppunct}\relax
\EndOfBibitem
\bibitem[Evans and Morriss(1983)Evans, and Morriss]{eva832}
Evans,~D.~J.; Morriss,~G. \emph{Phys. Lett. A} \textbf{1983}, \emph{98},
  433--436\relax
\mciteBstWouldAddEndPuncttrue
\mciteSetBstMidEndSepPunct{\mcitedefaultmidpunct}
{\mcitedefaultendpunct}{\mcitedefaultseppunct}\relax
\EndOfBibitem
\bibitem[Minary \latin{et~al.}(2003)Minary, Martyna, and Tuckerman]{min031}
Minary,~P.; Martyna,~G.~J.; Tuckerman,~M.~E. \emph{J. Chem. Phys.}
  \textbf{2003}, \emph{118}, 2510--2526\relax
\mciteBstWouldAddEndPuncttrue
\mciteSetBstMidEndSepPunct{\mcitedefaultmidpunct}
{\mcitedefaultendpunct}{\mcitedefaultseppunct}\relax
\EndOfBibitem
\bibitem[Collins \latin{et~al.}(2010)Collins, Ezra, and Wiggins]{col101}
Collins,~P.; Ezra,~G.~S.; Wiggins,~S. \emph{J. Chem. Phys.} \textbf{2010},
  \emph{133}, 014105\relax
\mciteBstWouldAddEndPuncttrue
\mciteSetBstMidEndSepPunct{\mcitedefaultmidpunct}
{\mcitedefaultendpunct}{\mcitedefaultseppunct}\relax
\EndOfBibitem
\bibitem[Morishita(2000)]{mor001}
Morishita,~T. \emph{J. Chem. Phys.} \textbf{2000}, \emph{113}, 2976--2982\relax
\mciteBstWouldAddEndPuncttrue
\mciteSetBstMidEndSepPunct{\mcitedefaultmidpunct}
{\mcitedefaultendpunct}{\mcitedefaultseppunct}\relax
\EndOfBibitem
\bibitem[Morishita(2003)]{mor031}
Morishita,~T. \emph{J. Chem. Phys.} \textbf{2003}, \emph{119}, 7075--7082\relax
\mciteBstWouldAddEndPuncttrue
\mciteSetBstMidEndSepPunct{\mcitedefaultmidpunct}
{\mcitedefaultendpunct}{\mcitedefaultseppunct}\relax
\EndOfBibitem
\bibitem[Lemak and Balabaev(1994)Lemak, and Balabaev]{lem941}
Lemak,~A.~S.; Balabaev,~N.~K. \emph{Mol. Simul.} \textbf{1994}, \emph{13},
  177--187\relax
\mciteBstWouldAddEndPuncttrue
\mciteSetBstMidEndSepPunct{\mcitedefaultmidpunct}
{\mcitedefaultendpunct}{\mcitedefaultseppunct}\relax
\EndOfBibitem
\bibitem[Harvey \latin{et~al.}(1998)Harvey, Tan, and Cheatham]{har981}
Harvey,~S.~C.; Tan,~R. K.-Z.; Cheatham,~T.~E. \emph{J. Comput. Chem.}
  \textbf{1998}, \emph{19}, 726--740\relax
\mciteBstWouldAddEndPuncttrue
\mciteSetBstMidEndSepPunct{\mcitedefaultmidpunct}
{\mcitedefaultendpunct}{\mcitedefaultseppunct}\relax
\EndOfBibitem
\bibitem[Callen(1985)]{cal851}
Callen,~H.~B. \emph{Thermodynamics and an Introduction to Thermostatistics},
  2nd ed.; John Wiley \& Sons, 1985\relax
\mciteBstWouldAddEndPuncttrue
\mciteSetBstMidEndSepPunct{\mcitedefaultmidpunct}
{\mcitedefaultendpunct}{\mcitedefaultseppunct}\relax
\EndOfBibitem
\bibitem[{\c{C}}a{\u{g}}in and Ray(1988){\c{C}}a{\u{g}}in, and Ray]{cag881}
{\c{C}}a{\u{g}}in,~T.; Ray,~J.~R. \emph{Phys. Rev. A} \textbf{1988}, \emph{37},
  4510--4513\relax
\mciteBstWouldAddEndPuncttrue
\mciteSetBstMidEndSepPunct{\mcitedefaultmidpunct}
{\mcitedefaultendpunct}{\mcitedefaultseppunct}\relax
\EndOfBibitem
\bibitem[Shirts \latin{et~al.}(2006)Shirts, Burt, and Johnson]{shi061}
Shirts,~R.~B.; Burt,~S.~R.; Johnson,~A.~M. \emph{J. Chem. Phys.} \textbf{2006},
  \emph{125}, 164102\relax
\mciteBstWouldAddEndPuncttrue
\mciteSetBstMidEndSepPunct{\mcitedefaultmidpunct}
{\mcitedefaultendpunct}{\mcitedefaultseppunct}\relax
\EndOfBibitem
\bibitem[Uline \latin{et~al.}(2008)Uline, Siderius, and Corti]{uli081}
Uline,~M.~J.; Siderius,~D.~W.; Corti,~D.~S. \emph{J. Chem. Phys.}
  \textbf{2008}, \emph{128}, 124301\relax
\mciteBstWouldAddEndPuncttrue
\mciteSetBstMidEndSepPunct{\mcitedefaultmidpunct}
{\mcitedefaultendpunct}{\mcitedefaultseppunct}\relax
\EndOfBibitem
\bibitem[Siboni \latin{et~al.}(2013)Siboni, Raabe, and Varnik]{sib131}
Siboni,~N.~H.; Raabe,~D.; Varnik,~F. \emph{Phys. Rev. E} \textbf{2013},
  \emph{87}, 030101\relax
\mciteBstWouldAddEndPuncttrue
\mciteSetBstMidEndSepPunct{\mcitedefaultmidpunct}
{\mcitedefaultendpunct}{\mcitedefaultseppunct}\relax
\EndOfBibitem
\bibitem[Cooke and Schmidler(2008)Cooke, and Schmidler]{coo081}
Cooke,~B.; Schmidler,~S.~C. \emph{J. Chem. Phys.} \textbf{2008}, \emph{129},
  164112\relax
\mciteBstWouldAddEndPuncttrue
\mciteSetBstMidEndSepPunct{\mcitedefaultmidpunct}
{\mcitedefaultendpunct}{\mcitedefaultseppunct}\relax
\EndOfBibitem
\bibitem[Lingenheil \latin{et~al.}(2008)Lingenheil, Denschlag, Reichold, and
  Tavan]{lin081}
Lingenheil,~M.; Denschlag,~R.; Reichold,~R.; Tavan,~P. \emph{J. Chem. Theory
  Comput.} \textbf{2008}, \emph{4}, 1293--1306\relax
\mciteBstWouldAddEndPuncttrue
\mciteSetBstMidEndSepPunct{\mcitedefaultmidpunct}
{\mcitedefaultendpunct}{\mcitedefaultseppunct}\relax
\EndOfBibitem
\bibitem[Goga \latin{et~al.}(2012)Goga, Rzepiela, de~Vries, Marrink, and
  Berendsen]{gog121}
Goga,~N.; Rzepiela,~A.~J.; de~Vries,~A.~H.; Marrink,~S.~J.; Berendsen,~H. J.~C.
  \emph{J. Chem. Theory Comput.} \textbf{2012}, \emph{8}, 3637--3649\relax
\mciteBstWouldAddEndPuncttrue
\mciteSetBstMidEndSepPunct{\mcitedefaultmidpunct}
{\mcitedefaultendpunct}{\mcitedefaultseppunct}\relax
\EndOfBibitem
\bibitem[Basconi and Shirts(2013)Basconi, and Shirts]{bas131}
Basconi,~J.~E.; Shirts,~M.~R. \emph{J. Chem. Theory Comput.} \textbf{2013},
  \emph{9}, 2887--2899\relax
\mciteBstWouldAddEndPuncttrue
\mciteSetBstMidEndSepPunct{\mcitedefaultmidpunct}
{\mcitedefaultendpunct}{\mcitedefaultseppunct}\relax
\EndOfBibitem
\bibitem[Nos{\'e}(1991)]{nos911}
Nos{\'e},~S. \emph{Prog. Theor. Phys. Suppl.} \textbf{1991}, \emph{103},
  1--46\relax
\mciteBstWouldAddEndPuncttrue
\mciteSetBstMidEndSepPunct{\mcitedefaultmidpunct}
{\mcitedefaultendpunct}{\mcitedefaultseppunct}\relax
\EndOfBibitem
\bibitem[Chiu \latin{et~al.}(2000)Chiu, Clark, Subramaniam, and
  Jakobsson]{chi001}
Chiu,~S.; Clark,~M.; Subramaniam,~S.; Jakobsson,~E. \emph{J. Comput. Chem.}
  \textbf{2000}, \emph{21}, 121--131\relax
\mciteBstWouldAddEndPuncttrue
\mciteSetBstMidEndSepPunct{\mcitedefaultmidpunct}
{\mcitedefaultendpunct}{\mcitedefaultseppunct}\relax
\EndOfBibitem
\bibitem[Eastwood \latin{et~al.}(2010)Eastwood, Stafford, Lippert, Jensen,
  Maragakis, Predescu, Dror, and Shaw]{eas101}
Eastwood,~M.~P.; Stafford,~K.~A.; Lippert,~R.~A.; Jensen,~M.~{\O}.;
  Maragakis,~P.; Predescu,~C.; Dror,~R.~O.; Shaw,~D.~E. \emph{J. Chem. Theory
  Comput.} \textbf{2010}, \emph{6}, 2045--2058\relax
\mciteBstWouldAddEndPuncttrue
\mciteSetBstMidEndSepPunct{\mcitedefaultmidpunct}
{\mcitedefaultendpunct}{\mcitedefaultseppunct}\relax
\EndOfBibitem
\bibitem[Sagui and Darden(1999)Sagui, and Darden]{sag991}
Sagui,~C.; Darden,~T.~A. \emph{Annu. Rev. Biophys. Biomol. Struct.}
  \textbf{1999}, \emph{28}, 155--179\relax
\mciteBstWouldAddEndPuncttrue
\mciteSetBstMidEndSepPunct{\mcitedefaultmidpunct}
{\mcitedefaultendpunct}{\mcitedefaultseppunct}\relax
\EndOfBibitem
\bibitem[Wagner \latin{et~al.}(2013)Wagner, Balaraman, Niesen, Larsen, Jain,
  and Vaidehi]{wag131}
Wagner,~J.~R.; Balaraman,~G.~S.; Niesen,~M. J.~M.; Larsen,~A.~B.; Jain,~A.;
  Vaidehi,~N. \emph{J. Comput. Chem.} \textbf{2013}, \emph{34}, 904--914\relax
\mciteBstWouldAddEndPuncttrue
\mciteSetBstMidEndSepPunct{\mcitedefaultmidpunct}
{\mcitedefaultendpunct}{\mcitedefaultseppunct}\relax
\EndOfBibitem
\bibitem[Yan \latin{et~al.}(2013)Yan, Sun, and Liu]{yan132}
Yan,~L.; Sun,~C.; Liu,~H. \emph{Adv. Manuf.} \textbf{2013}, \emph{1},
  160--165\relax
\mciteBstWouldAddEndPuncttrue
\mciteSetBstMidEndSepPunct{\mcitedefaultmidpunct}
{\mcitedefaultendpunct}{\mcitedefaultseppunct}\relax
\EndOfBibitem
\bibitem[Plimpton(1995)]{pli951}
Plimpton,~S. \emph{J. Comput. Phys.} \textbf{1995}, \emph{117}, 1--19\relax
\mciteBstWouldAddEndPuncttrue
\mciteSetBstMidEndSepPunct{\mcitedefaultmidpunct}
{\mcitedefaultendpunct}{\mcitedefaultseppunct}\relax
\EndOfBibitem
\bibitem[Not()]{Note-1}
We used the \DTMdate{2016-11-17} release of LAMMPS to conduct our simulations.
  The Gaussian thermostat was not implemented in LAMMPS, so we wrote an
  extension that integrates the equations of motion given by \citet{min031}.
  This extension was later incorporated into the LAMMPS code and made publicly
  available starting with the \DTMdate{2017-01-06} update as part of the ``fix
  nvk'' command.\relax
\mciteBstWouldAddEndPunctfalse
\mciteSetBstMidEndSepPunct{\mcitedefaultmidpunct}
{}{\mcitedefaultseppunct}\relax
\EndOfBibitem
\bibitem[Martin and Siepmann(1998)Martin, and Siepmann]{mar982}
Martin,~G.~M.; Siepmann,~J.~I. \emph{J. Phys. Chem. B} \textbf{1998},
  \emph{102}, 2569--2577\relax
\mciteBstWouldAddEndPuncttrue
\mciteSetBstMidEndSepPunct{\mcitedefaultmidpunct}
{\mcitedefaultendpunct}{\mcitedefaultseppunct}\relax
\EndOfBibitem
\bibitem[Toxvaerd and Olsen(1990)Toxvaerd, and Olsen]{tox901}
Toxvaerd,~S.; Olsen,~O.~H. \emph{Phys. Scr.} \textbf{1990}, \emph{1990},
  98--101\relax
\mciteBstWouldAddEndPuncttrue
\mciteSetBstMidEndSepPunct{\mcitedefaultmidpunct}
{\mcitedefaultendpunct}{\mcitedefaultseppunct}\relax
\EndOfBibitem
\bibitem[Tuckerman \latin{et~al.}(2001)Tuckerman, Liu, Ciccotti, and
  Martyna]{tuc011}
Tuckerman,~M.~E.; Liu,~Y.; Ciccotti,~G.; Martyna,~G.~J. \emph{J. Chem. Phys.}
  \textbf{2001}, \emph{115}, 1678--1702\relax
\mciteBstWouldAddEndPuncttrue
\mciteSetBstMidEndSepPunct{\mcitedefaultmidpunct}
{\mcitedefaultendpunct}{\mcitedefaultseppunct}\relax
\EndOfBibitem
\bibitem[Hess(2003)]{hes031}
Hess,~S. \emph{Z. Naturforsch. A} \textbf{2003}, \emph{58}, 377--391\relax
\mciteBstWouldAddEndPuncttrue
\mciteSetBstMidEndSepPunct{\mcitedefaultmidpunct}
{\mcitedefaultendpunct}{\mcitedefaultseppunct}\relax
\EndOfBibitem
\bibitem[Manousiouthakis and Deem(1999)Manousiouthakis, and Deem]{man991}
Manousiouthakis,~V.~I.; Deem,~M.~W. \emph{J. Chem. Phys.} \textbf{1999},
  \emph{110}, 2753--2756\relax
\mciteBstWouldAddEndPuncttrue
\mciteSetBstMidEndSepPunct{\mcitedefaultmidpunct}
{\mcitedefaultendpunct}{\mcitedefaultseppunct}\relax
\EndOfBibitem
\bibitem[Mark and Nilsson(2002)Mark, and Nilsson]{mar021}
Mark,~P.; Nilsson,~L. \emph{J. Comput. Chem.} \textbf{2002}, \emph{23},
  1211--1219\relax
\mciteBstWouldAddEndPuncttrue
\mciteSetBstMidEndSepPunct{\mcitedefaultmidpunct}
{\mcitedefaultendpunct}{\mcitedefaultseppunct}\relax
\EndOfBibitem
\bibitem[Mudi and Chakravarty(2004)Mudi, and Chakravarty]{mud041}
Mudi,~A.; Chakravarty,~C. \emph{Mol. Phys.} \textbf{2004}, \emph{102},
  681--685\relax
\mciteBstWouldAddEndPuncttrue
\mciteSetBstMidEndSepPunct{\mcitedefaultmidpunct}
{\mcitedefaultendpunct}{\mcitedefaultseppunct}\relax
\EndOfBibitem
\bibitem[Berendsen \latin{et~al.}(1981)Berendsen, Postma, van Gunsteren, and
  Hermans]{ber811}
Berendsen,~H. J.~C.; Postma,~J. P.~M.; van Gunsteren,~W.~F.; Hermans,~J.
  Interaction Models for Water in Relation to Protein Hydration. Intermolecular
  Forces. Dordrecht, 1981; pp 331--342\relax
\mciteBstWouldAddEndPuncttrue
\mciteSetBstMidEndSepPunct{\mcitedefaultmidpunct}
{\mcitedefaultendpunct}{\mcitedefaultseppunct}\relax
\EndOfBibitem
\bibitem[Steele(1973)]{ste731}
Steele,~W.~A. \emph{Surf. Sci.} \textbf{1973}, \emph{36}, 317--352\relax
\mciteBstWouldAddEndPuncttrue
\mciteSetBstMidEndSepPunct{\mcitedefaultmidpunct}
{\mcitedefaultendpunct}{\mcitedefaultseppunct}\relax
\EndOfBibitem
\bibitem[Chandler(1987)]{cha871}
Chandler,~D. \emph{Introduction to Modern Statistical Mechanics}; Oxford
  University Press, 1987\relax
\mciteBstWouldAddEndPuncttrue
\mciteSetBstMidEndSepPunct{\mcitedefaultmidpunct}
{\mcitedefaultendpunct}{\mcitedefaultseppunct}\relax
\EndOfBibitem
\bibitem[Shirts(2013)]{shi131}
Shirts,~M.~R. \emph{J. Chem. Theory Comput.} \textbf{2013}, \emph{9},
  909--926\relax
\mciteBstWouldAddEndPuncttrue
\mciteSetBstMidEndSepPunct{\mcitedefaultmidpunct}
{\mcitedefaultendpunct}{\mcitedefaultseppunct}\relax
\EndOfBibitem
\bibitem[Cheng and Merz(1996)Cheng, and Merz]{che961}
Cheng,~A.; Merz,~K.~M. \emph{J. Phys. Chem.} \textbf{1996}, \emph{100},
  1927--1937\relax
\mciteBstWouldAddEndPuncttrue
\mciteSetBstMidEndSepPunct{\mcitedefaultmidpunct}
{\mcitedefaultendpunct}{\mcitedefaultseppunct}\relax
\EndOfBibitem
\bibitem[Mor \latin{et~al.}(2008)Mor, Ziv, and Levy]{mor081}
Mor,~A.; Ziv,~G.; Levy,~Y. \emph{J. Comput. Chem.} \textbf{2008}, \emph{29},
  1992--1998\relax
\mciteBstWouldAddEndPuncttrue
\mciteSetBstMidEndSepPunct{\mcitedefaultmidpunct}
{\mcitedefaultendpunct}{\mcitedefaultseppunct}\relax
\EndOfBibitem
\bibitem[Rosta \latin{et~al.}(2009)Rosta, Buchete, and Hummer]{ros091}
Rosta,~E.; Buchete,~N.~V.; Hummer,~G. \emph{J. Chem. Theory Comput.}
  \textbf{2009}, \emph{5}, 1393--1399\relax
\mciteBstWouldAddEndPuncttrue
\mciteSetBstMidEndSepPunct{\mcitedefaultmidpunct}
{\mcitedefaultendpunct}{\mcitedefaultseppunct}\relax
\EndOfBibitem
\bibitem[Spill \latin{et~al.}(2011)Spill, Pasquali, and Derreumaux]{spi111}
Spill,~Y.~G.; Pasquali,~S.; Derreumaux,~P. \emph{J. Chem. Theory Comput.}
  \textbf{2011}, \emph{7}, 1502--1510\relax
\mciteBstWouldAddEndPuncttrue
\mciteSetBstMidEndSepPunct{\mcitedefaultmidpunct}
{\mcitedefaultendpunct}{\mcitedefaultseppunct}\relax
\EndOfBibitem
\bibitem[Bussi and Parrinello(2008)Bussi, and Parrinello]{bus081}
Bussi,~G.; Parrinello,~M. \emph{Comput. Phys. Commun.} \textbf{2008},
  \emph{179}, 26--29\relax
\mciteBstWouldAddEndPuncttrue
\mciteSetBstMidEndSepPunct{\mcitedefaultmidpunct}
{\mcitedefaultendpunct}{\mcitedefaultseppunct}\relax
\EndOfBibitem
\bibitem[Tafipolsky \latin{et~al.}(2007)Tafipolsky, Amirjalayer, and
  Schmid]{taf071}
Tafipolsky,~M.; Amirjalayer,~S.; Schmid,~R. \emph{J. Comput. Chem.}
  \textbf{2007}, \emph{28}, 1169--1176\relax
\mciteBstWouldAddEndPuncttrue
\mciteSetBstMidEndSepPunct{\mcitedefaultmidpunct}
{\mcitedefaultendpunct}{\mcitedefaultseppunct}\relax
\EndOfBibitem
\bibitem[Amirjalayer \latin{et~al.}(2007)Amirjalayer, Tafipolsky, and
  Schmid]{ami071}
Amirjalayer,~S.; Tafipolsky,~M.; Schmid,~R. \emph{Angew. Chem., Int. Ed.}
  \textbf{2007}, \emph{46}, 463--466\relax
\mciteBstWouldAddEndPuncttrue
\mciteSetBstMidEndSepPunct{\mcitedefaultmidpunct}
{\mcitedefaultendpunct}{\mcitedefaultseppunct}\relax
\EndOfBibitem
\bibitem[Smit and Maesen(2008)Smit, and Maesen]{smi081}
Smit,~B.; Maesen,~T. L.~M. \emph{Chem. Rev.} \textbf{2008}, \emph{108},
  4125--4184\relax
\mciteBstWouldAddEndPuncttrue
\mciteSetBstMidEndSepPunct{\mcitedefaultmidpunct}
{\mcitedefaultendpunct}{\mcitedefaultseppunct}\relax
\EndOfBibitem
\bibitem[Seehamart \latin{et~al.}(2009)Seehamart, Nanok, Krishna, van Baten,
  Remsungnen, and Fritzsche]{see091}
Seehamart,~K.; Nanok,~T.; Krishna,~R.; van Baten,~J.~M.; Remsungnen,~T.;
  Fritzsche,~S. \emph{Microporous Mesoporous Mater.} \textbf{2009}, \emph{125},
  97--100\relax
\mciteBstWouldAddEndPuncttrue
\mciteSetBstMidEndSepPunct{\mcitedefaultmidpunct}
{\mcitedefaultendpunct}{\mcitedefaultseppunct}\relax
\EndOfBibitem
\bibitem[Dubbeldam \latin{et~al.}(2007)Dubbeldam, Walton, Ellis, and
  Snurr]{dub071}
Dubbeldam,~D.; Walton,~K.~S.; Ellis,~D.~E.; Snurr,~R.~Q. \emph{Angew. Chem.
  Int. Ed.} \textbf{2007}, \emph{46}, 4496--4499\relax
\mciteBstWouldAddEndPuncttrue
\mciteSetBstMidEndSepPunct{\mcitedefaultmidpunct}
{\mcitedefaultendpunct}{\mcitedefaultseppunct}\relax
\EndOfBibitem
\bibitem[Ford \latin{et~al.}(2009)Ford, Dubbeldam, and Snurr]{for091}
Ford,~D.~C.; Dubbeldam,~D.; Snurr,~R.~Q. \emph{diffusion-fundamentals.org}
  \textbf{2009}, \emph{11}, 1--8\relax
\mciteBstWouldAddEndPuncttrue
\mciteSetBstMidEndSepPunct{\mcitedefaultmidpunct}
{\mcitedefaultendpunct}{\mcitedefaultseppunct}\relax
\EndOfBibitem
\bibitem[Ponder and Richards(1987)Ponder, and Richards]{pon871}
Ponder,~J.~W.; Richards,~F.~M. \emph{J. Comput. Chem.} \textbf{1987}, \emph{8},
  1016--1024\relax
\mciteBstWouldAddEndPuncttrue
\mciteSetBstMidEndSepPunct{\mcitedefaultmidpunct}
{\mcitedefaultendpunct}{\mcitedefaultseppunct}\relax
\EndOfBibitem
\bibitem[Braun \latin{et~al.}(2015)Braun, Chen, Schnell, Lin, Reimer, and
  Smit]{bra152}
Braun,~E.; Chen,~J.~J.; Schnell,~S.~K.; Lin,~L.-C.; Reimer,~J.~A.; Smit,~B.
  \emph{Angew. Chem. Int. Ed.} \textbf{2015}, \emph{54}, 14349--14352\relax
\mciteBstWouldAddEndPuncttrue
\mciteSetBstMidEndSepPunct{\mcitedefaultmidpunct}
{\mcitedefaultendpunct}{\mcitedefaultseppunct}\relax
\EndOfBibitem
\bibitem[Leyssale and Vignoles(2008)Leyssale, and Vignoles]{ley081}
Leyssale,~J.-M.; Vignoles,~G.~L. \emph{Chem. Phys. Lett.} \textbf{2008},
  \emph{454}, 299--304\relax
\mciteBstWouldAddEndPuncttrue
\mciteSetBstMidEndSepPunct{\mcitedefaultmidpunct}
{\mcitedefaultendpunct}{\mcitedefaultseppunct}\relax
\EndOfBibitem
\bibitem[Wong-ekkabut \latin{et~al.}(2010)Wong-ekkabut, Miettinen, Dias, and
  Karttunen]{won101}
Wong-ekkabut,~J.; Miettinen,~M.~S.; Dias,~C.; Karttunen,~M. \emph{Nat.
  Nanotechnol.} \textbf{2010}, \emph{5}, 555--557\relax
\mciteBstWouldAddEndPuncttrue
\mciteSetBstMidEndSepPunct{\mcitedefaultmidpunct}
{\mcitedefaultendpunct}{\mcitedefaultseppunct}\relax
\EndOfBibitem
\bibitem[Baker(2016)]{bak161}
Baker,~M. \emph{Nature} \textbf{2016}, \emph{533}, 452--454\relax
\mciteBstWouldAddEndPuncttrue
\mciteSetBstMidEndSepPunct{\mcitedefaultmidpunct}
{\mcitedefaultendpunct}{\mcitedefaultseppunct}\relax
\EndOfBibitem
\bibitem[Hu and Sinnott(2004)Hu, and Sinnott]{hu041}
Hu,~Y.; Sinnott,~S.~B. \emph{J. Comput. Phys.} \textbf{2004}, \emph{200},
  251--266\relax
\mciteBstWouldAddEndPuncttrue
\mciteSetBstMidEndSepPunct{\mcitedefaultmidpunct}
{\mcitedefaultendpunct}{\mcitedefaultseppunct}\relax
\EndOfBibitem
\end{mcitethebibliography}

\clearpage
\appendix
\appendixpage

\section{Equipartition in the isokinetic ensemble}

To the best of our knowledge, it has not been shown that the
equipartition theorem need necessarily apply in the isokinetic ensemble,
and it is not immediately clear that it must.  When additional
constraints are added to the system, such as the constraint of a
constant \gls{com} momentum that is typical in \gls{md} simulations with
\gls{pbc} or the constraint of a constant kinetic energy in the
isokinetic ensemble, the change to the partition function can bring
about a changed type of energy partitioning.\cite{shi061,uli081}

To illustrate, we can briefly examine the former constraint of constant
\gls{com} momentum, which has been analyzed before.\cite{cag881,shi061,uli081}
One might naively think that the equipartition theorem for \acrlongpl{dof}
related to the constraint (in this case, kinetic \acrlongpl{dof}, $p_i$)
would simply change to:
\begin{equation}
\label{eq:equipartitionmomentumconstrainednaive}
    \left<H_{p_i}\right>
    = \frac{1}{2}k_{\text{B}} T \frac{N-1}{N}
\end{equation}
However, this is incorrect. Instead, it can be shown that for the
canonical ensemble with its \gls{com} momentum constrained to zero, the
principle of energy equipartitioning is violated for kinetic
\acrlongpl{dof}.\cite{uli081} The system instead obeys the equation:
\begin{equation}
    \left<H_{p,i}\right>
    = \frac{1}{2}k_{\text{B}} T \frac{M_{\text{total}}-m_i}{M_{\text{total}}}
\end{equation}
with a similar expression for the microcanonical ensemble when the
\gls{com} momentum is constrained to zero.\cite{uli081} For a system of
equally-sized particles, the naive expression of
Eq.~\ref{eq:equipartitionmomentumconstrainednaive} is recovered and
equipartition holds, but for a system with particles of difference
masses, the particles will have different amounts of kinetic energy; for
a system containing massive tracer particles, the difference between the
expressions can be severe.\cite{sib131}
In the thermodynamic limit, the constraint of constant \gls{com}
momentum does not affect the equipartition theorem.

Our equipartition theorem analysis of the isokinetic ensemble very
closely follows the work of \citet{uli081} for the momentum-constrained
canonical ensemble. The system to be analyzed is described by the
Hamiltonian:
\begin{equation}
    H(\mathbf{r}^N, \mathbf{p}^N) = \sum_{i=1}^N
    \frac{\mathbf{p}_i^2}{2m_i} + U(\mathbf{r}^N)
\end{equation}

The configurational part of the isokinetic ensemble's partition function
is not interesting since it is equivalent to that of the canonical
ensemble's. We will focus on the integral over momenta, or the
``translational'' partition function of the isokinetic
ensemble:\cite{min031,uli081}
\begin{equation}
    Q_{\text{trans}}(N,V,T,K) = \int d\mathbf{p}^N \exp \left( -\beta
    \sum_{i=1}^N \frac{\mathbf{p}_i^2}{2m_i} \right) \delta
    \left( K - \sum_{i=1}^N \frac{\mathbf{p}_i^2}{2m_i} \right)
\end{equation}
To solve this expression, we will take the Laplace transform, integrate,
and then take the inverse Laplace transform.  We Laplace transform
$Q_{\text{trans}}$ from $K \rightarrow s$ to obtain:
\begin{equation}
    Q_{\text{trans}}(N,V,T,s) = \int d\mathbf{p}^N \exp \left( -\left(
    \beta + s \right) \sum_{i=1}^N \frac{\mathbf{p}_i^2}{2m_i} \right)
\end{equation}
Integrating over all $\mathbf{p}_i$ gives:
\begin{equation}
    Q_{\text{trans}}(N,V,T,s) = \left( 2\pi \right)^{\frac{dN}{2}}
    \left( \prod_{i=1}^N m_i^{\frac{d}{2}} \right) \left( s+\beta
    \right)^{-\frac{dN}{2}}
\end{equation}
The inverse Laplace transform from $s \rightarrow K$ yields:
\begin{equation}
    Q_{\text{trans}}(N,V,T,K) = \left( 2\pi \right)^{\frac{dN}{2}}
    \left( \prod_{i=1}^N m_i^{\frac{d}{2}} \right) \frac{\exp (-\beta K)
    K^{\frac{dN}{2}-1}}{\Gamma \left( \frac{dN}{2} \right)}
\end{equation}

This translational partition function is then used to generate the
probability distribution $f$ for a single kinetic \acrlong{dof}, $p_1$:
\begin{equation}
    f(N,V,T,K,p_1) = \frac {\int dp^{dN-1} \exp \left(
        -\beta \sum_{i=1}^{dN} \frac{p_i^2}{2m_i} \right)
    \delta \left( K - \sum_{i=1}^{dN} \frac{p_i^2}{2m_i} \right)}
    {Q_{\text{trans}}(N,V,T,K)}
\end{equation}
As before, we Laplace transform ($K \rightarrow s$), integrate, and
inverse Laplace transform ($s \rightarrow K$) to obtain:
\begin{align}
    f(N,V,T,K,p_1)
    &= \frac {\left( 2\pi \right)^{\frac{dN-1}{2}} \frac{\left(
    \prod_{i=1}^N m_i^{\frac{d}{2}} \right)} {m_1^{\frac{1}{2}}}
    \frac{\exp (-\beta K) \left( K-\frac{p_1^2}{2m_1} \right)
    ^{\frac{dN-1}{2}-1} \Theta \left( K-\frac{p_1^2}{2m_1} \right)}
    {\Gamma \left(\frac{dN-1}{2} \right)}} {Q_{\text{trans}}(N,V,T,K)}
    \nonumber \\
    &=\left( 2\pi m_1 \right)^{-\frac{1}{2}} \frac{\left(
    K-\frac{p_1^2}{2m_1} \right)^{\frac{dN-1}{2}-1}}{K^{\frac{dN}{2}-1}}
    \frac{\Gamma \left( \frac{dN}{2} \right)}{\Gamma \left(
    \frac{dN-1}{2} \right)} \Theta \left( K-\frac{p_1^2}{2m_1} \right)
\end{align}
The presence of the Heaviside step function is a consequence of the
impossibility of satisfying the kinetic energy constraint if the kinetic
energy of a single \acrlong{dof} is greater than the set total kinetic
energy. The function can be integrated over $p_1$ by setting the
integration bounds as $p_1=\pm\sqrt{2m_1K}$ to remove the Heaviside step
function from the integral.  It can be verified that the integral of
$f(N,V,T,K,p_1)$ over $p_1$ is unity.

The average kinetic energy of a kinetic \acrlong{dof} is then:
\begin{align}
    \frac{<p_1^2>}{2m_1}
    &= \int dp_1 \frac{p_1^2}{2m_1} f(N,V,T,K,p_1) \nonumber \\
    &= \frac{K}{dN}
\end{align}
which indicates equipartition for every kinetic \acrlong{dof}, regardless of
the value of $K$. If $K$ is set to the average kinetic energy for a
particular temperature, i.e., $K=\frac{d}{2} N k_{\text{B}} T$:
\begin{equation}
    \frac{<p_1^2>}{2m_1} = \frac{1}{2}k_{\text{B}} T
\end{equation}
so every \acrlong{dof} will have the same average kinetic energy as in
the canonical or microcanonical ensembles.

Thus, the equipartition theorem must hold in the isokinetic ensemble.

\section{Partitioning kinetic energy into translational, rotational,
and vibrational modes}

Kinetic energies of each diatomic molecule were partitioned into translational,
rotational, and vibrational kinetic energies. In one dimension, this was
done as:
\begin{align}
    K
    &= \frac{1}{2}m_1v_{1,x}^2+\frac{1}{2}m_2v_{2,x}^2 \nonumber \\
    &= \frac{1}{2}\left(m_1+m_2\right)\left(\frac{m_1v_{1,x}+m_2v_{2,x}}{m_1+m_2}\right)^2
     + \frac{1}{2}\left(\frac{m_1m_2}{m_1+m_2}\right)\left(v_{2,x}-v_{1,x}\right)^2 \nonumber \\
    &= \underbrace{\frac{1}{2}\left(m_1+m_2\right)v_{\text{trans},x}^2}_{K_{\rm{trans}}}
     + \underbrace{\frac{1}{2}\mu v_{\text{vib},x}^2}_{K_{\rm{vib}}} \nonumber
\end{align}
where $v_{\text{trans},i}=\frac{m_1v_{1,i}+m_2v_{2,i}}{m_1+m_2}$,
$v_{\text{vib},i}=v_{2,i}-v_{1,i}$, and
$\mu=\frac{m_1m_2}{m_1+m_2}$, giving one translational and one
vibrational \acrlongpl{dof} for the molecule.  In three dimensions,
this was similarly done as:
\begin{align}
    K
    &= \frac{1}{2}m_1\left(v_{1,x}^2+v_{1,y}^2+v_{1,z}^2\right)
     +\frac{1}{2}m_2\left(v_{2,x}^2 + v_{2,y}^2+v_{2,z}^2\right) \nonumber \\
    &= \underbrace{\frac{1}{2}\left(m_1+m_2\right)\left(v_{\text{trans},x}^2
     + v_{\text{trans},y}^2+v_{\text{trans},z}^2\right)}_{K_{\rm{trans}}}
     + \underbrace{\frac{1}{2}\mu\left(\left(v_{2,x}-v_{1,x}\right)^2 + \left(v_{2,y}-v_{1,y}\right)^2\right)}_{K_{\rm{rot}}}
     + \underbrace{\frac{1}{2}\mu\left(v_{2,z}-v_{1,z}\right)^2}_{K_{\rm{vib}}} \nonumber
\end{align}
with an arbitrary coordinate axis aligned with the bond vector (we chose
to label the equation above such that the $z$-axis was aligned), giving
three translational, two rotational, and one vibrational \acrlongpl{dof} for
the molecule. The code to calculate these partitioned energies was
incorporated into the open-source LAMMPS code and made publicly
available starting with the \DTMdate{2016-09-13} update as part of the
``compute bond/local'' command.

\section{Simulation details for the benzene in MOF-5 system}

Simulations of benzene in \acrshort{mof}-5 were conducted with the Tinker
package,\cite{pon871} version 7.1, for the purposes of using the force
field of \citet{taf071} to compare results with \citet{ami071}.  Tinker
input scripts are available in the Supplementary Information\dag. The
force field of \citet{taf071} was used with the modification of using
point charges instead of bond dipoles since---to the best of our
knowledge---computing Ewald summations with the latter is not
implemented in Tinker (it is unclear how Ewald summations were computed
in \citet{taf071} and \citet{ami071}); we used the atomic charges that
\citet{taf071} used to parameterize their force field, as given in their
Table~2.  We strived to keep conditions as similar to those used by
\citet{ami071} as possible; we used a timestep of \SI{1}{\femto\second},
a Lennard-Jones potential cutoff of \SI{12}{\angstrom}, Ewald summations
with default Tinker 7.1 parameters, the formerly default Berendsen
thermostat time damping constant of \SI{100}{\femto\second} (except for
where we noted that we used the currently default time damping constant
of \SI{200}{\femto\second}), and \gls{nosehoover} default Tinker
parameters. Simulations were run for at least \SI{2}{\nano\second} of
equilibration and at least \SI{100}{\nano\second} of production, which
was found to be sufficiently long for the \gls{msd} to become a linear
function of time.  Simulations were conducted in a simulation box
consisting of a single cubic unit cell taken from a minimized structure
described in \citet{taf071} and kindly supplied to us by Rochus Schmid,
consisting of 424 atoms and with a lattice constant of
\SI{25.9457}{\angstrom}. The 10 benzene molecules were distributed
through the \acrshort{mof}-5 crystal by running a \SI{200}{\pico\second}
\gls{md} simulation at \SI{1000}{\kelvin} with the \gls{mof} atoms
frozen prior to equilibration.

\section{Derivation of phase space boundaries in Fig.~\ref{fig:balanceviolation}}

Consider a \gls{md} simulation initially on line $\overline{CD}$ in
phase space (Fig.~\ref{fig:balanceviolation}). By being infinitesimally close to point $C$ when
rescaling, the rescaling line will have a slope of
$\frac{K_{\rm{target}}-x_C}{x_C}$. To remain moving in phase space in
the direction of increasing translational kinetic energy, rescaling must
continue to occur below the target isokinetic line; thus, the greatest
slope that can continue to be achieved is $\frac{K_{\rm{target}}-x}{x}$.
We see that the red dashed line can therefore be derived by solving the
differential equation
$\frac{dy_{\rm{max}}}{dx}=\frac{K_{\rm{target}}-x}{x}$ with boundary
condition $y_{\rm{max}}(x_C)=y_C=K_{\rm{target}}-x_C$, which results in
$y_{\rm{max}}(x)=K_{\rm{target}}\ln{\left(\frac{x}{x_C}\right)}-x+K_{\rm{target}}$.

A similar situation occurs when moving in phase space in the direction
of decreasing translational kinetic energy. The velocity rescaling line
with smallest slope is achieved when rescaling just above the target
isokinetic line, e.g., just above point $F$ when rescaling from line
$\overline{AB}$. The same differential equation must be solved to derive
the blue dashed line, only changing the boundary condition to
$y_{\rm{max}}(x_B)=y_B$, which results in
$y_{\rm{max}}(x)=K_{\rm{target}}\ln{\left(\frac{x}{x_B}\right)}-x+x_B+y_B$.
By replacing $x_B$ and $y_B$ in this equation with the beginning
positions $x_0$ and $y_0=U_{\rm{max}}(x_0)$, one obtains the general expression
$y_{\rm{max}}(x)=K_{\rm{target}}\ln{\left(\frac{x}{x_0}\right)}-x+x_0+U_{\rm{max}}(x_0)$.

\clearpage
\section{Supplementary Figures}

\begin{figure}[h]
  \centering
  \includegraphics[width=3.34in]{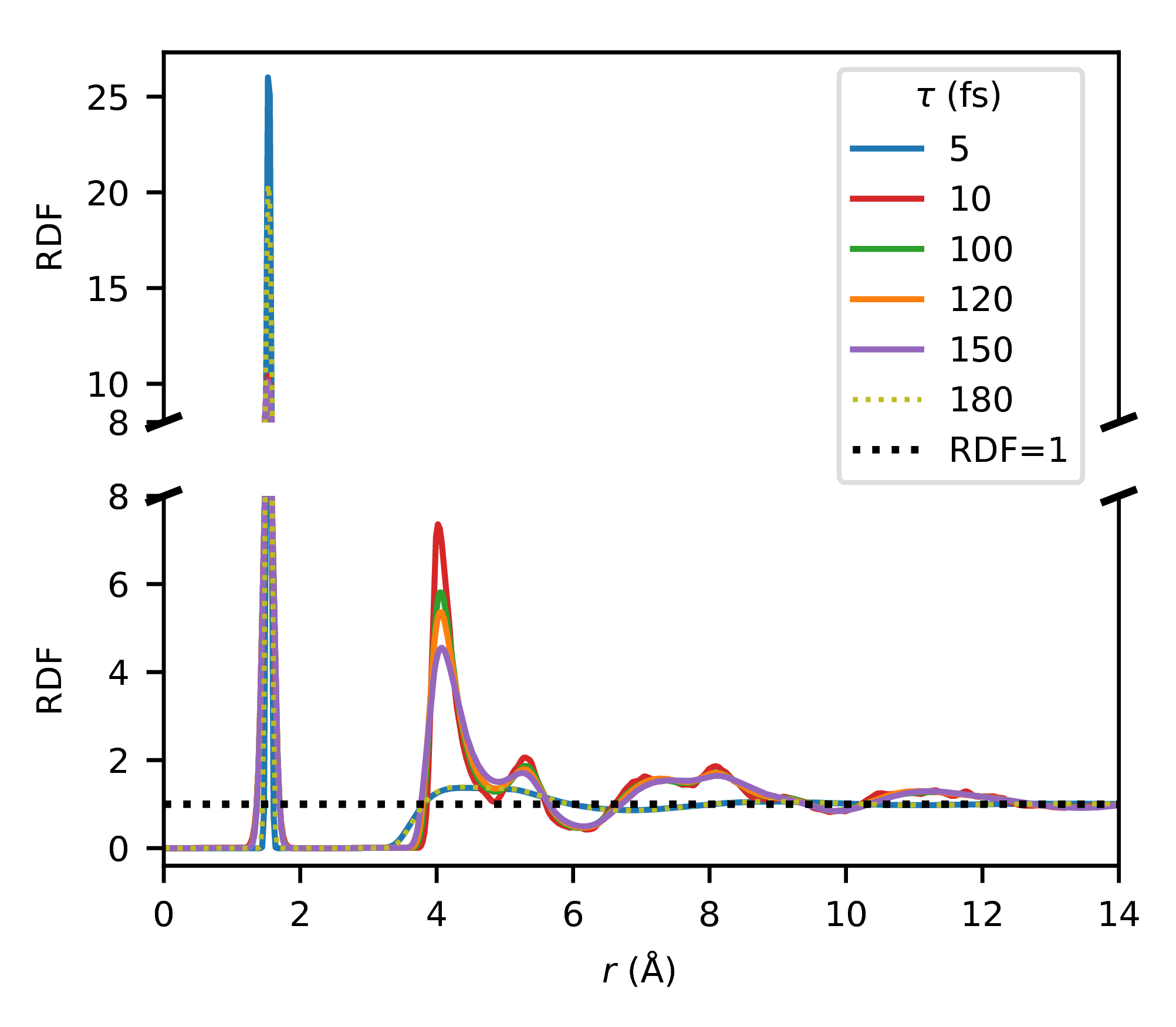}
  \caption{\label{fig:rdfSI}
    Similar to Fig.~\ref{fig:rdf}, \acrfull{rdf} of the CH$_3$-CH$_3$ distance obtained
    from \gls{md} simulations of 235 ethane molecules in a
    \SI{30}{\angstrom} cubic simulation box with a target temperature
    set to \SI{256}{\kelvin} using the Berendsen thermostat with
    different values of the time damping constant.  \glspl{rdf} for
    simulations with the time damping constant set to
    \SIlist{0.5;2}{\femto\second} look similar to the
    \SI{5}{\femto\second} case, and \glspl{rdf} for simulations with the
    time damping constant set to \SIlist{200;1000}{\femto\second} look
    similar to the \SI{180}{\femto\second} case.}
\end{figure}

\begin{figure}[h]
  \centering
  \includegraphics[width=3.48in]{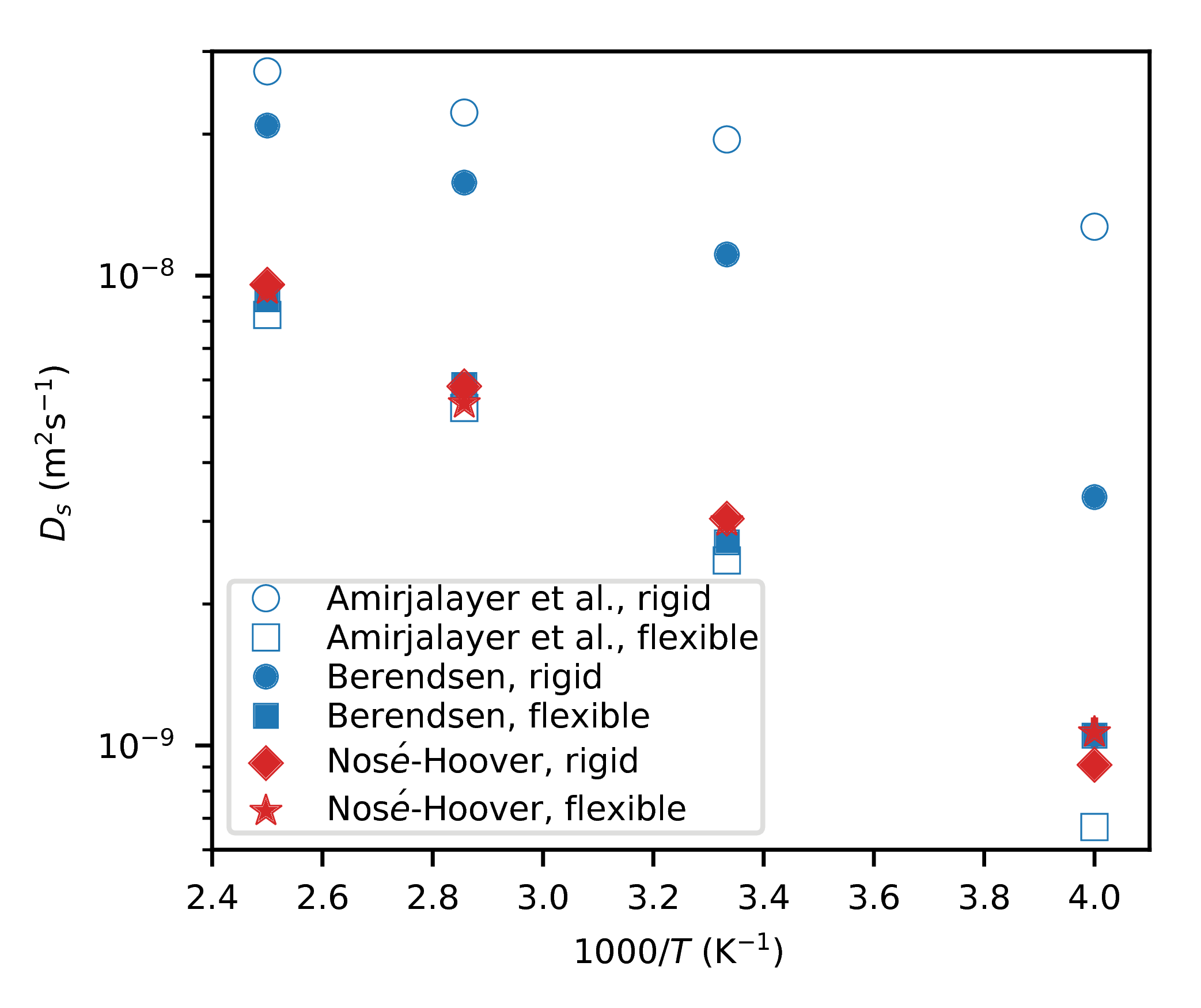}
  \caption{\label{fig:diffcoeffSI}
    Identical to Fig.~\ref{fig:diffcoeff}, except the Berendsen thermostat was used with a time
    damping constant of \SI{200}{\femto\second} instead of
    \SI{100}{\femto\second}.}
\end{figure}

\end{singlespace}
\end{document}